\documentclass[11pt,a4paper]{article}

\usepackage{lmodern} % readable font in pdf
\usepackage[utf8]{inputenc} % input special characters
\usepackage[T1]{fontenc} % output special characters
\usepackage{microtype} % better font spacing
\usepackage[english]{babel}

\usepackage{color,graphicx}

\usepackage{amsmath,dsfont,url,amssymb}
\usepackage{slashed,cancel}
\usepackage[usenames,dvipsnames]{xcolor}
\usepackage[colorlinks=true, linkcolor=black, citecolor=ForestGreen]{hyperref}%BrickRed
\usepackage{mdframed}
\usepackage{overpic,subfigure}

\newcommand{\be}{\begin{equation}}
\newcommand{\ee}{\end{equation}}
\newcommand\p{\ensuremath{\partial}}

\definecolor{inkred}{RGB}{210,29,0}
\definecolor{inkblue}{RGB}{0,112,196}

\def \d {\partial}

\DeclareMathOperator{\Tr}{Tr}

\numberwithin{equation}{section}

\begin{document}
\frenchspacing

\title{
\begin{flushright}\vspace{-1.1in}
			\mbox{\normalsize  EFI-21-1}
		\end{flushright} 
		\vspace{40pt}
Large Charge Sector of 3d Parity-Violating CFTs \\[0.5em]}

\author{Gabriel Cuomo$^{a,b}$, Luca V. Delacr\'etaz$^c$ and Umang Mehta$^c$ \\[0.5em]
{
\small
 \it ${}^a$Simons Center for Geometry and Physics, Stony Brook University, Stony Brook, NY 11794, USA}\\ 
 \small
{ \it ${}^b$C. N. Yang Institute for Theoretical Physics, Stony Brook University, Stony Brook, NY 11794, USA}\\
\small
{\it ${}^c$Kadanoff Center for Theoretical Physics, University of Chicago, Chicago, IL 60637, USA}\\[1em]
}

\date{}
\maketitle

\begin{abstract}

Certain CFTs with a global $U(1)$ symmetry become superfluids when coupled to a chemical potential. When this happens, a Goldstone effective field theory controls the spectrum and correlators of the lightest large charge operators. We show that in 3d, this EFT contains a single parity-violating 1-derivative term with quantized coefficient. This term forces the superfluid ground state to have vortices on the sphere, leading to a spectrum of large charge operators that is remarkably richer than in parity-invariant CFTs. We test our predictions in a weakly coupled Chern-Simons matter theory.

\end{abstract}

\pagebreak
\tableofcontents
\pagebreak

%######################################################################%
%======================================================================%
%======================================================================%
%======================================================================%
%######################################################################%
\section{Introduction and Results}

Universality, or the stringent constraint of scale invariance and unitarity, implies that a number of quantum field theories and other many body systems are described by the same Conformal Field Theory (CFT) when tuned to a fixed point. Strikingly, different CFTs themselves share `super-universal' features, such as the spectrum and correlators of large spin \cite{Alday:2007mf,Fitzpatrick:2012yx,Komargodski:2012ek,Caron-Huot:2017vep}, large charge \cite{Hellerman:2015nra,Monin:2016jmo,Jafferis:2017zna,Hellerman:2017sur} and heavy \cite{Lashkari:2016vgj,Alday:2019qrf,Delacretaz:2020nit,Belin:2020hea} operators. Some of these features can be established and studied in the absence of a small parameter in the underlying CFT.

In CFTs with an internal global symmetry, operators of large charge $Q\gg 1$ under that symmetry can be probed by turning on a chemical potential $\mu$. Doing so can drive the CFT into a number of phases, including a superfluid, a Landau Fermi liquid, an extremal black hole, or possibly a non-Fermi liquid. In any of these examples the chemical potential sources both a finite energy and charge density in the thermodynamic limit%
	\footnote{There are also situations where this does not happen, such as a free complex scalar, or in theories with a moduli space.}%
, implying a scaling of the lightest operators at fixed charge $\Delta_{\rm min}(Q)\sim Q^{d/(d-1)}$ in $d$ spacetime dimensions. 

In this paper, following Refs.~\cite{Hellerman:2015nra,Monin:2016jmo} we study the large charge sector of CFTs in $d=3$ with a global $U(1)$ symmetry, assuming they enter a superfluid phase. Superfluids are described by a local effective field theory (EFT), which provides a simple tool to make controlled predictions about the large charge spectrum of the underlying CFT. We focus on CFTs that do not have parity (or time-reversal) symmetry. In this case, we find that the EFT contains a single parity-violating 1-derivative correction. It is best formulated in dual language in terms of a gauge field $a_\mu$, and takes the form
\begin{equation}\label{eq_EFT_intro}
\begin{split}
S_{\rm EFT}[a]
	&= -\alpha \int d^3x \,  |f|^{3/2} \\
	&+ \frac{\kappa}{8\pi} \int d^3 x \sqrt{-g} \, a_\mu \epsilon^{\mu\nu\lambda} \epsilon_{\alpha\beta\gamma} u^\alpha \left(\nabla_\nu u^\beta \nabla_\lambda u^\gamma - \frac12 \mathcal R_{\nu\lambda}{}^{\beta\gamma}\right) \\
	& + O(\d^2) \, , 
\end{split}
\end{equation}
where $f_{\mu\nu} = \d_\mu a_\nu - \d_\nu a_\mu$, and $u_\mu$ is a unit vector proportional to $\epsilon_{\mu\nu\lambda}f^{\nu\lambda}$. The leading term with coefficient $\alpha$ is the superfluid stiffness and preserves parity; it controls the trajectory of the lightest large charge operators $\Delta_{\rm min}(Q) = \sqrt{2} \pi \alpha Q^{3/2} + O(Q^{1/2})$. The second term has a quantized coefficient $\kappa\in \mathbb Z$, and was first introduced in Ref.~\cite{Golkar:2014paa}. It is $O(\d)$ and hence suppressed compared to the stiffness, but more relevant than the $O(\d^2)$ parity-preserving higher-derivative corrections \cite{Monin:2016jmo}. Despite its derivative suppression, it has crucial consequences for the spectrum of lightest operators at fixed large charge, which are described as finite density states on the sphere $\mathbb R\times S^2$. 
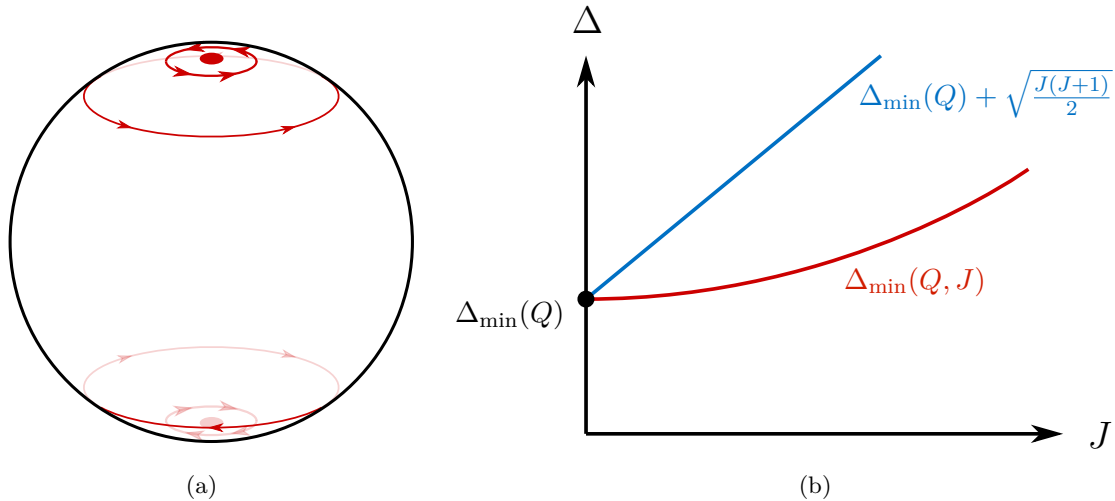
\begin{figure}
\vspace{10pt}
\centerline{
\hspace{-20pt}
\subfigure[]{
\includegraphics[width=0.35\linewidth, angle=0]{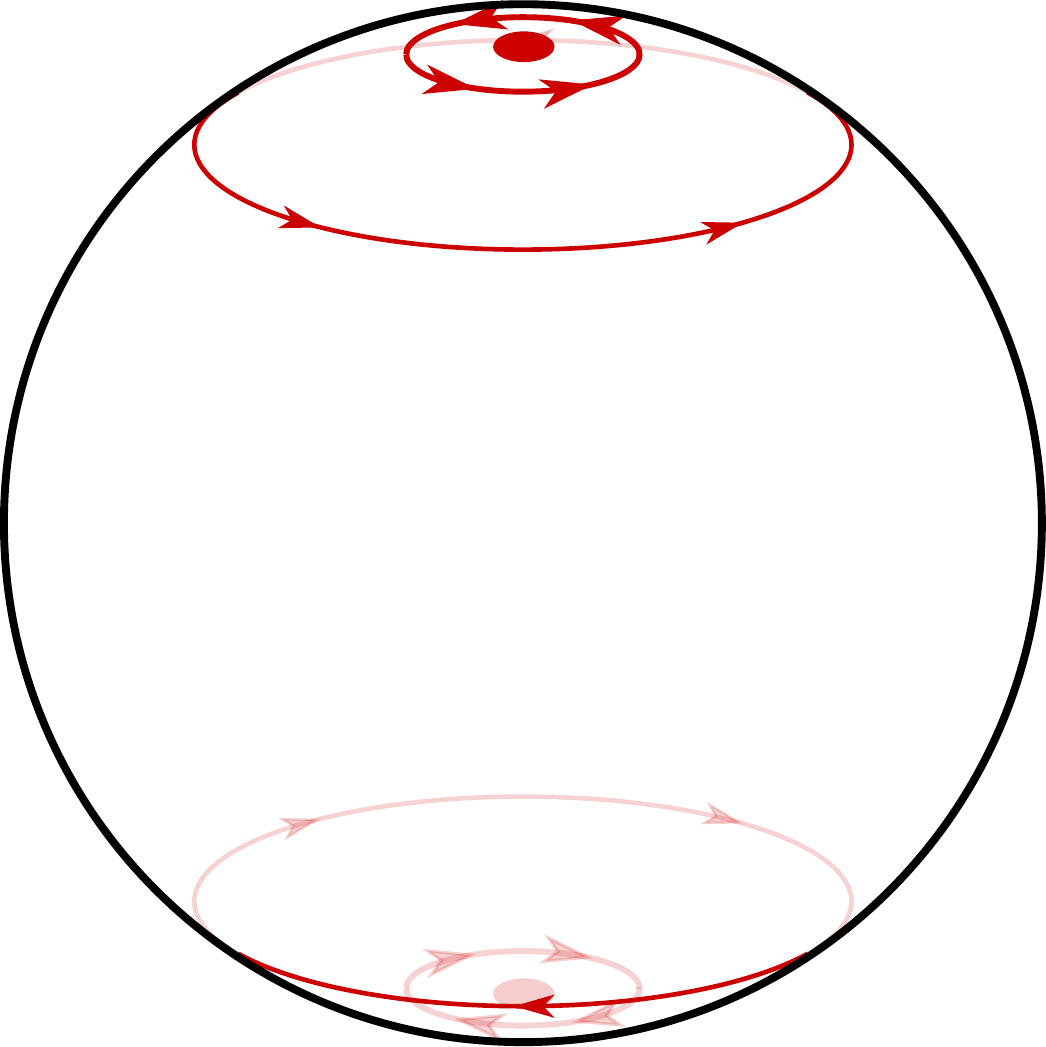}}
\hspace{50pt}
\subfigure[]{
\begin{overpic}[width=0.42\textwidth,tics=10]{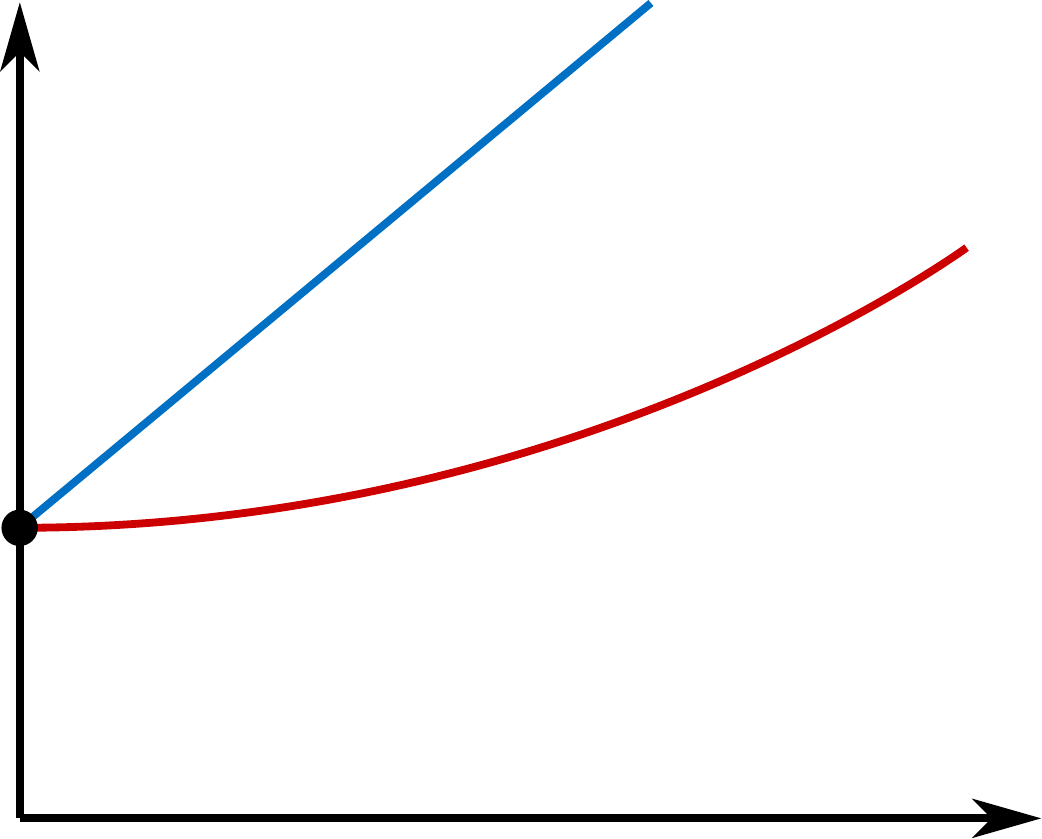}
	 \put (-1,85) {\Large$\Delta$} 
	 \put (-25,25) {$\Delta_{\rm min}(Q)$} 
	 \put (55,32) {{\color{inkred}$\Delta_{\rm min}(Q,J)$}} 
	 \put (58,70) {{\color{inkblue}$\Delta_{\rm min}(Q)+\sqrt{\frac{J(J+1)}{2}}$}} 
	 \put (105,-1) {\Large$J$} 
\end{overpic}
}}
\caption{\label{fig_kappa2} Classically, the superfluid ground state has $\kappa$ vortices arranged to minimize their potential energy, shown in (a) for $\kappa=2$. Quantum mechanically, the true ground state has vanishing spin but vortex excitations are soft. For $\kappa=2$, the spectrum of vortex excitations is given in Eq.~\eqref{eq_DeltaQJ_intro} and shown in (b) in red with the spectrum of single phonon states in blue. 
}
\end{figure}
We show that the EFT \eqref{eq_EFT_intro} can only be consistently placed on the sphere with vortices, with total vorticity $\kappa$ (in the dual picture \eqref{eq_EFT_intro}, the new term introduces a background charge density on the sphere which must be neutralized by charged particles). Classically, the ground state is then described by $|\kappa|$ vortices in a configuration which minimizes their Coulomb energy -- an extremization problem known as Whyte's problem in the mathematical literature. Although the solution is not known for general $\kappa$, it can be shown to have vanishing angular momentum when $|\kappa|\neq1$, so that the spin of the corresponding lightest operator vanishes. When $|\kappa|=1$, the ground state contains a single vortex with angular momentum $J=Q/2$ -- the lightest operators of large charge $Q$ in CFTs with $|\kappa|=1$ therefore also have large spin $J=Q/2$. In all of these cases, the vortex fugacities give a contribution to the dimension of the lightest large charge operators
\begin{equation}\label{eq_DeltaQ_intro}
\Delta_{\rm min}(Q) = 
	\sqrt{2}\pi \alpha Q^{3/2} + \frac{\kappa }{12\sqrt{2}\pi\alpha} \sqrt{Q}\log Q  + O(\sqrt{Q})\, .
	%\frac{2}{3\sqrt{2\pi \chi_0}} \, Q^{3/2} + \kappa \frac{\sqrt{2\pi \chi_0}}{8} \sqrt{Q}\log \frac{Q}{\chi_0} + O(\sqrt{Q})\, .
\end{equation}
Quantizing the system leads to a rich low-lying spectrum of vortex excitations, which are softer than superfluid phonons and hence describe the lightest operators of large charge and finite spin $0<J\lesssim \sqrt{Q}$. For example, when $\kappa=2$ one finds 
\begin{equation}\label{eq_DeltaQJ_intro}
\Delta_{\rm min}(Q,J) \simeq
	\Delta_{\rm min}(Q) + \frac{1}{6\sqrt{2}\pi \alpha} \frac{J(J+1)}{Q^{3/2}}\, .
\end{equation}
The spectrum of large charge states is illustrated in Fig.~\ref{fig_kappa2}. Although the spectrum in CFTs with $|\kappa|>2$ is more complicated, Eq.~\eqref{eq_DeltaQJ_intro} still holds parametrically, i.e. the lightest operators still satisfy $\Delta_{\rm min}(Q,J) - \Delta_{\rm min}(Q) \sim J^2/Q^{3/2}$. Quantization also shows that although the vortex configuration is inhomogeneous classically, in the true quantum ground state the vortices are delocalized.

These results should hold in any 3d parity-violating CFT that becomes a superfluid at finite density. We show that these predictions are borne out in a weakly coupled Chern-Simons matter theory, consisting of a single Dirac fermion coupled to a dynamical gauge field
\begin{equation}\label{eq_S_anyon_intro}
S_{\rm anyon\, CFT}
	= \int d^3x\, \bar\psi i\cancel D \psi - \frac{k}{4\pi} \epsilon^{\mu\nu\lambda} a_\mu\d_\nu a_\lambda\, , 
\end{equation}
with level $k\gg 1$. A non-relativistic cousin of this theory goes under the name of anyon superfluid \cite{PhysRevLett.60.2677,Chen:1989xs}, and we show that the CFT \eqref{eq_S_anyon_intro} also becomes a superfluid at finite density. Weak coupling $\sim 1/k$ of this CFT allows one to derive the corresponding superfluid EFT \eqref{eq_EFT_intro} directly from \eqref{eq_S_anyon_intro}, thereby obtaining expressions for the EFT parameters $\alpha$ and $\kappa$. Large charge operators in this theory are monopoles, dressed with fermions occupying the first $k$ Landau levels. The degeneracy of Landau levels on the sphere (or monopole harmonics) implies that $\kappa$ fermions are missing from the $k$th Landau level -- these are the $\kappa$ vortices expected from the EFT. 

The rest of this paper is organized as follows: the EFT \eqref{eq_EFT_intro} for parity-violating conformal superfluids is constructed and studied classically in Sec.~\ref{sec_EFT}. It is quantized in Sec.~\ref{sec_CFTdata}, where the spectrum of lightest operators at large charge is discussed, and Eqs.~\eqref{eq_DeltaQ_intro}, \eqref{eq_DeltaQJ_intro} are obtained. As the leading parity-violating effect in the EFT, $\kappa$ also gives the leading contribution to parity-odd heavy-heavy-light OPE coefficients in the large charge limit, which are also discussed in Sec.~\ref{sec_CFTdata}.  Finally, we test these predictions in the weakly coupled CFT \eqref{eq_S_anyon_intro} and discuss further applications in Sec.~\ref{sec_appli}.

%######################################################################%
%======================================================================%
%======================================================================%
%======================================================================%
%######################################################################%
\section{EFT of conformal superfluids}\label{sec_EFT}

%We are interested in studying a CFT at finite density, or chemical potential $\mu$. There are a number of phases that can describe the finite density CFT, including a superfluid, a Fermi liquid or an extremal black hole. Note that only the first is described by a strictly local EFT.

%======================================================================%
%======================================================================%
%======================================================================%
\subsection{Review of parity-preserving EFT}

We briefly review the construction of the EFT for conformal superfluids, referring the reader to Ref.~\cite{Monin:2016jmo} for details. In a superfluid phase, a Goldstone $\phi$ nonlinearly realizes the internal $U(1)$ symmetry (as well as certain spacetime symmetries); the most general invariant Lagrangian may be easily obtained requiring $U(1)$ and Weyl invariance.
%coset construction can be used to generate all invariant the most general effective action, in a gradient expansion. 
Retaining terms up to second order in derivatives, one obtains:
\begin{align}\nonumber
S =&-\frac{c_1}{d(d-1)}\int d^dx\sqrt{g} |\p\phi|^d \\ \label{eq_chiEFT}
&+c_2\int d^dx\sqrt{g} |\p\phi|^d\left\{ \frac{\mathcal{R}}{|\p\phi|^2}
+(d-1)(d-2)\frac{\left[\nabla_\mu|\p\phi|\right]^2}{|\p\phi|^4}\right\}\\
\nonumber
&+c_3\int d^dx\sqrt{g} |\p\phi|^d\left\{\mathcal{R}_{\mu\nu}\frac{\p^\mu\phi\p^\nu\phi}{|\p\phi|^4}
+(d-1)(d-2)\frac{\left[\p^\mu\phi\nabla_\mu|\p\phi|\right]^2}{|\p\phi|^6}
\right. \\
\nonumber
&\hspace*{3.5cm} \;\;\left.
+(d-2)\nabla_\mu\left[\frac{\p^\mu\phi\p^\nu\phi}{|\p\phi|^2}\right]
\frac{\nabla_\nu|\p\phi|}{|\p\phi|^3}
\right\}+\cdots
% +O\left(|\p\phi|^{d}\frac{\nabla^4}{|\p\phi|^{4}}\right) \,,
\end{align}
where $\mathcal R_{\mu\nu}$ and $\mathcal R \equiv \mathcal R_\mu^\mu$ are the Ricci tensor and scalar for the background metric $g$.
We will only need the leading term with coefficient $c_1$ in this paper. The terms 
beyond the first line are two-derivative suppressed compared to the leading term, and $\cdots$ denotes higher derivative terms. In the next section, we will see that the leading parity-violating terms are only one-derivative suppressed compared to $c_1$.

The microscopic $U(1)$ symmetry is nonlinearly realized on the scalar field as $\phi \to \phi + c$. In addition, the superfluid phase has an emergent ($d-2$)-form symmetry $U(1)^{(d-2)}$ that counts the winding of $\phi$. The currents for both of these symmetries are
\begin{equation}\label{eq_currentschi}
j_\mu = \frac{c_1}{d-1}|\d\phi|^{d-2}\d_\mu \phi + \cdots \, , \qquad\qquad
J_{\mu_1 \cdots \mu_{d-1}}
	= \frac{1}{2\pi} \epsilon_{\mu_1 \cdots \mu_d} \d^{\mu_d} \phi\, ,
\end{equation}
where $\cdots$ denotes higher derivative terms. The currents are normalized such that their integrals over closed manifolds produce integer valued charges $\int_{M_d}\star j$ and $\int_{M_1} \star J$.

%======================================================================%
%======================================================================%
%======================================================================%
\subsection{Parity-violating terms}

Parity-violation is better studied after dualizing, namely replacing the scalar degree of freedom $\phi$ with a gauge field (see e.g.~\cite{Cuomo:2017vzg} where this is done in a similar context). This can be achieved by replacing $\d_\mu \phi \to v_\mu$ and introducing a Lagrangian parameter $\frac{i}{2\pi}\int a \wedge d v$ that forces $v_\mu$ to be longitudinal. Integrating $v_\mu$ out then leads to the effective action
\begin{equation}\label{eq_alphaEFT}
S = -\alpha \int d^{d} x \sqrt{-g} |f|^{\frac{d}{d-1}} + \cdots
\end{equation}
with $f_\mu\equiv \epsilon_{\mu\lambda_1 \cdots \lambda_{d-1}}\d^{\lambda_1} a^{\lambda_2 \cdots \lambda_{d-1}}$ and $|f| \equiv \sqrt{-f_\mu f^\mu}$, and $\cdots$ denotes higher derivative terms coming e.g.~from $c_2$ and $c_3$ in \eqref{eq_chiEFT}. The coefficient $\alpha$ is given in terms of $c_1$ by Eq.~\eqref{eq_susceptibility} below. The currents \eqref{eq_currentschi} are now given by
\begin{equation}\label{eq_currents_a}
j_\mu = \frac{1}{2\pi} f_\mu \, , 
%j_\mu = \frac{1}{2\pi} \epsilon_{\mu \mu_1\cdots \mu_d } \d^{\mu_1} a^{\mu_2\cdots \mu_d} \, , 
\qquad\qquad
J_{\mu_1 \cdots \mu_{d-1}}
	%= \frac{1}{2\pi} \epsilon_{\mu_1 \cdots \mu_d} \left( \frac{d-1}{2\pi\chi_0}\right)^{\frac1{d-1}} \frac{f^{\mu_d}}{|f|^{\frac{d-2}{d-1}}}
	= \alpha \frac{d}{d-1}\epsilon_{\mu_1 \cdots \mu_d} \frac{f^{\mu_d}}{|f|^{\frac{d-2}{d-1}}} \, .
\end{equation}
and the stress tensor is
\begin{equation}\label{eq_T_original}
T_{\mu\nu} = \frac{d}{d-1} \alpha |f|^{\frac{d}{d-1}} \left(u_\mu u_\nu + \frac{1}{d}g_{\mu\nu}\right) + \cdots \, ,
\end{equation}
where we have introduced a unit vector $u_\mu \equiv f_\mu/|f|$ satisfying $u_\mu u^\mu=-1$. Around finite density backgrounds $\langle j^\mu\rangle = \frac{1}{2\pi} \langle f^\mu\rangle = \rho \delta^\mu_0$, the energy density is $\varepsilon = \alpha (2\pi \rho)^{\frac{d}{d-1}}$, which leads to the relation $\Delta_{\rm min}(Q) \sim \alpha Q^{\frac{d}{d-1}}$ for large charge operators studied in Sec.~\ref{sec_CFTdata}. The dimensionless charge susceptibility of the superfluid is given by
\begin{equation}\label{eq_susceptibility}
\chi_0 \equiv \frac{1}{\mu^{d-1}} \frac{d \rho}{d\mu} = c_1 = \frac{d-1}{(2\pi)^{d}} \left(\frac{d-1}{d\,\alpha} \right)^{d-1}.
\end{equation}
We will interchangeably use $\chi_0$ or $\alpha$ in the rest of the paper.

We now specialize to $d = 3$ spacetime dimensions. In this dual picture it was found in Ref.~\cite{Golkar:2014paa} that one can write two parity-violating terms consistent with symmetries that are only one-derivative suppressed compared to the leading term \eqref{eq_alphaEFT}%
	\footnote{The action \eqref{eq_aEFT} can presumably also be obtained directly using an appropriate coset construction, along the lines of Ref.~\cite{Goon:2014ika}. One advantage of this somewhat more tedious approach is that all terms can be obtained `algorithmically', including the $\kappa$ term which arises as a Wess-Zumino term following Refs.~\cite{Goon:2012dy,Delacretaz:2014jka}.}
\begin{equation}\label{eq_aEFT}
\begin{split}
S
	&= - \alpha \int d^3x \sqrt{-g}\,  |f|^{3/2} \\
	&+ \zeta \int d^3 x\, |f| \epsilon^{\mu\nu\lambda} u_\mu\d_\nu u_\lambda\\
	&+ \frac{\kappa}{8\pi} \int d^3 x \sqrt{-g} \, a_\mu \epsilon^{\mu\nu\lambda} \epsilon_{\alpha\beta\gamma} u^\alpha \left(\nabla_\nu u^\beta \nabla_\lambda u^\gamma - \frac12 \mathcal R_{\nu\lambda}{}^{\beta\gamma}\right) \\
	& + O(\d^2) \, , 
\end{split}
\end{equation}
We are focusing here on parity-violating superfluids with conformal symmetry, the more general case is discussed in Appendix \ref{app_EFT}. The two new terms $\zeta,\, \kappa$ are less relevant than the leading term $\alpha$, but more relevant than the subleading  corrections $c_2,\,c_3$ in \eqref{eq_chiEFT}. The advantage of working in the dual picture is now clear: the $\zeta$ term would vanish if written in terms of $u_\mu = \d_\mu\chi / |\d\chi|$, and the $\kappa$ term would be non-local%
	\footnote{Said differently, the $\kappa$ term explicitly breaks the emergent winding (1-form) symmetry, which acts on the gauge field as $a_\mu \to a_\mu + c_\mu$. Non-conservation of the 1-form current will have crucial consequences in Sec.~\ref{ssec_EFTsphere}. 
	%This symmetry is only broken by nonperturbative effects in \eqref{eq_chiEFT}.
	}.
Note that this term is gauge invariant because $a_\mu$ is contracted with an identically conserved current
\begin{equation}\label{eq_JEuler}
J_{\rm Euler}^\mu 
	\equiv \frac{1}{8\pi} \epsilon^{\mu\nu\lambda} \epsilon_{\alpha\beta\gamma} u^\alpha \left(\nabla_\nu u^\beta \nabla_\lambda u^\gamma - \frac12 \mathcal R_{\nu\lambda}{}^{\beta\gamma}\right) \, , 
\end{equation}
called `Euler current' in Ref.~\cite{Golkar:2014paa}. The name stems from the fact its integral on a spatial manifold is proportional to its Euler charateristic -- see e.g.~Eq.~\eqref{eq_eulerdensity} below. Invariance under large gauge transformations requires
\begin{equation}\label{eq_quant}
\kappa \in \mathbb Z\, .
\end{equation}
{We do not include a Chern-Simons term in \eqref{eq_aEFT}; as a relevant term breaking the 1-form symmetry, it would gap the system. For generic values of the Wilson coefficients, the superfluid phonon would receive a mass of the order of the EFT cutoff%
	\footnote{In Sec.~\ref{ssec_anyon}, we apply the EFT to a weakly coupled CFT, where the presence of a small parameter allows the inclusion of a Chern-Simons term without producing an empty theory below the cutoff (see in particular Eq.~\eqref{eq_SEFTprime}).}. Since the coefficient of the Chern-Simons term is quantized and does not receive radiative corrections, choosing its coefficient to vanish is technically natural.}

In the conformal context, the EFT \eqref{eq_aEFT} can in fact be simplified. Since the $\zeta$ term vanishes on the leading equations of motion (because of conservation of $J_{\mu\nu}$ in \eqref{eq_currents_a}), one might expect that it can be removed with a field redefinition. Although this is not quite correct in the general case (cf.~Appendix \ref{app_EFT}), it is for the conformal superfluid where the field redefinition is
\begin{equation}\label{eq_fieldredef}
a_\mu \to a_\mu  + \frac{2\zeta}{3\alpha} \frac{f_\mu}{|f|^{1/2}}\, .
\end{equation}
Now this field redefinition does not preserve the normalization $\oint \frac{a}{2\pi}\in \mathbb Z$, unless $\zeta = \left(\frac{3\alpha}{2}\right)^2 n$ with $n\in \mathbb Z$. In this case \eqref{eq_fieldredef} reads
\begin{equation}\label{eq_a_redef_J}
a_\mu \to a_\mu + n (\star J)_\mu\, , 
\end{equation}
which preserves the normalization because of the quantization of the higher form charge \eqref{eq_currents_a}. The part of $\zeta$ that cannot be removed is therefore circle-valued $\zeta = \left(\frac{3\alpha}{2}\right)^2 \frac{\theta}{2\pi}$ with $\theta\in [0,2\pi)$ %%]$
and has similar properties to a $\theta$-term in gauge theory. The final EFT for parity-violating conformal superfluids hence takes the form
\begin{equation}\label{eq_aEFT_theta}
\begin{split}
S
	&= - \alpha \int d^3x \sqrt{-g}\,  |f|^{3/2} \\
	&+ \theta \int (\star J)\wedge d (\star J)  + \kappa \int d^3 x \sqrt{-g} \, a_\mu J^\mu_{\rm Euler}\\
	& + O(\d^2) \, , 
\end{split}
\end{equation}
The $\theta$ term is a total derivative and it does not contribute in perturbation theory around the homogeneous background. Nonetheless, it has simple nonperturbative effects; for example, it will split the energy of positive and negative winding vortices on the plane $\mathbb R^2$. Since we will be working on a compact spatial manifold, namely the sphere $S^2$, we can drop all total derivatives and simply work with the action \eqref{eq_EFT_intro}.

The stress tensor of the theory \eqref{eq_aEFT_theta} is given by
\begin{equation}\label{eq_T}
T^{\mu\nu}
	= \frac{3}{2}\varepsilon u^\mu u^\nu +  \tilde\eta  \left[u_\alpha \epsilon^{\alpha\beta \mu}\nabla_\beta u^{\nu} + (\mu \leftrightarrow \nu)\right] - \rm trace + \cdots \, .
\end{equation}
with $\varepsilon = \alpha |f|^{3/2}$ and $\tilde \eta = \frac{\kappa}{8\pi}|f|$. This expression fits into the general form a stress tensor takes in a parity-violating fluid \cite{Jensen:2011xb}; a more thorough comparison with the hydrodynamics is given in appendix \ref{app_EFT}, where details for obtaining the stress tensor from \eqref{eq_aEFT_theta} are also given. In hydrodynamics, $\tilde \eta$ is referred to as the Hall viscosity, because it describes parity-odd momentum transport in finite density states $\langle f^\mu\rangle = 2\pi \rho \delta^\mu_0$ through the Kubo formula (see e.g.~\cite{PhysRevB.86.245309})
\begin{equation}\label{eq_etatilde}
\tilde \eta = \lim_{\omega\to 0} \frac{1}{i\omega}G^R_{T_{xx}T_{xy}}(\omega,k=0)\, .
\end{equation}
Eq.~\eqref{eq_quant} implies that $\tilde \eta$ is quantized in units of $\rho/4$ in relativistic superfluids.

%======================================================================%
%======================================================================%
%======================================================================%
\subsection{Placing the EFT on the sphere requires vortices}\label{ssec_EFTsphere}

In order to study the large charge spectrum of local operators in CFTs, we place the EFT \eqref{eq_aEFT} on a spatial sphere $ S^2$. Strikingly, it turns out that, despite its gradient suppression, the leading parity-odd term $\kappa$ has an important qualitative effect when studying the theory on the sphere: it forbids a homogeneous finite density solution $\langle f^\mu\rangle = 2\pi \rho \delta^\mu_0$ \cite{Read:1999fn,Golkar:2014paa,Moroz:2015cft}. This can be seen from the equation of motion for $a_\mu$, which reads as a non-conservation of the 1-form current (or simply as charged electric matter in dual language)%
	\footnote{{We expect the existence of higher derivative $O(\d^3)$ terms in the EFT that also break the 1-form symmetry (perhaps related to the  gravitational Chern-Simons term, see Refs.~\cite{Gromov:2014gta,CAN2015752}). These will give contributions to the right-hand side of \eqref{eq_noncons}, which however vanish when evaluated on the background, and therefore do not affect the discussion in this section.}}
\begin{equation}\label{eq_noncons}
\nabla	_\mu J^{\mu\nu}
	= \kappa J^\nu_{\rm Euler}\, .
\end{equation}
Looking for a solution close to the homogeneous finite density profile $\langle f^\mu\rangle = 2\pi \rho \delta^\mu_0$, which is a solution of the leading order equation of motion, one finds that the right-hand side is proportional to the Euler density on the sphere 
\begin{equation}\label{eq_eulerdensity}
J^\nu_{\rm Euler} \simeq \delta^\nu_0 \frac{\mathcal R}{8\pi} = \delta^\nu_0 \frac{1}{4\pi R^2}\, .
\end{equation}
Eq.~\eqref{eq_noncons} then requires a velocity field with a constant vorticity on the sphere, which is not possible without producing vortices%
\footnote{This vorticity can also be neutralized with a background flux of magnetic field for the $U(1)$ symmetry. In this context $\kappa$ is related to the `shift' of superfluids and quantum Hall systems \cite{PhysRevLett.69.953,PhysRevB.84.085316,Golkar:2014wwa,Golkar:2014paa}. Since CFT operators map to states on the sphere without any fluxes, we do not turn on any background fields in the following. 
}.
In dual language, \eqref{eq_noncons} introduces a constant electric charge density, which on the sphere must be neutralized by charged matter for Gauss's law to hold. Although we have here expanded around the finite density background, these conclusions hold more generally, see Ref.~\cite{Golkar:2014paa}.

We therefore generalize our EFT to include vortices. The spacetime trajectories of the vortices can be parametrized with worldlines $X^\mu_p(\tau)$. The most general action compatible with the symmetries of the system reads:
\begin{equation}\label{eq_Stot}
S = S_{\rm superfluid} + 
	S_{\rm vortices}\, ,
\end{equation}
with $S_{\rm superfluid}$ given by \eqref{eq_aEFT} and \cite{Horn:2015zna,Cuomo:2017vzg}
\begin{equation}
\begin{split}
S_{\rm vortices}
	&= -\sum_p \left(w_p\int_{X_p} a  + \gamma_p \int_{X_p}|f|^{1/2} \,dX_p + \cdots\right) \\ 
	%&= \sum_p \left(w_p \int d\tau\, a_\mu \dot X^\mu_p + \gamma_p \int_{X_p}d\tau \sqrt{f} \sqrt{-g_{\mu\nu} \dot X^\mu \dot X^\nu }   + \cdots \right) \, , \\
	&= -\sum_p \int d\tau\, \left(w_p \, a_\mu \dot X^\mu_p + \gamma_p  |f|^{1/2} \sqrt{-g_{\mu\nu} \dot X^\mu \dot X^\nu }   + \cdots  \right)\, , \\
\end{split}
\end{equation}
where $w_p\in \mathbb Z$ are the windings and $\gamma_p$ the dimensionless tensions of the vortices $p=1,2,\ldots,N_{\rm vortices}$. Equation \eqref{eq_noncons} now instead reads 
\begin{equation}\label{eq_noncons2}
\nabla	_\mu J^{\mu\nu}
	= \kappa J^\nu_{\rm Euler} - j^\nu_{\rm vortex}\, .
\end{equation}

A natural derivative counting scheme in the superfluid and vortex system is $f \sim \d a\sim 1$ and $\d X \sim 1$. In this scheme the $\alpha$ term is $O(\d^0)$, the $\kappa$ term and Wilson line term are $O(\d)$ (note that $\delta^2(x-X_p)\sim \d^2$), and the contribution from the vortex velocities in the tension term is $O(\d^2)$. In this paper, we only study the leading effects of parity violation -- we therefore only keep terms up to $O(\d)$ and work with the action 
\begin{equation}\label{eq_Sfinal}
\begin{split}
S	&= -\alpha \int d^3 x \sqrt{-g} |f|^{3/2} \\
	&+ \int d^3 x \sqrt{-g}\, a_\mu \Bigl(\kappa J^\mu_{\rm Euler} - \sum_p w_p\delta^2(x-X_p) \dot X^\mu_p \Bigr)\\
	&+ O(\d^2)\, .
\end{split}
\end{equation}
Quantum mechanically, the spectrum of a point particle on the sphere in a monopole magnetic field consists of Landau levels \cite{Wu:1976ge}. Dropping the mass term for the vortex is equivalent to restricting the description to the lowest Landau level \cite{PhysRevD.41.661,DUNNE1993114}. Indeed, the gap separation of higher Landau levels is given by the cyclotron frequency formula $\rho/m_{\rm vortex}\sim\sqrt{\rho}/\gamma$, which coincides parametrically with the EFT cutoff.

We now can compute the energy and angular momentum deriving from the action \eqref{eq_Sfinal} for an arbitrary classical state; our analysis will be almost identical to the one in Ref.~\cite{Cuomo:2017vzg}. The results obtained here will then be used in the next section to study the CFT spectrum. 
We begin by expanding the fields around a solution with finite density $\rho$:
\begin{equation}\label{eq_finiterho}
\langle j^\mu\rangle = \frac{1}{2\pi} \langle f^\mu\rangle = \rho \delta^\mu_0\, ,
\end{equation}
and write the electric and magnetic fields as
\begin{equation}
f^0 = 2\pi \rho - b \, , \qquad f^i = -\epsilon^{ij} e_j\, .
\end{equation}
To first order in perturbation theory, the equations of motion reduce to the ones of electrostatic for point charges on the sphere in the presence of a homogeneous charge density:
\begin{align} \label{eq_LorentzF}
\alpha \frac{3/2}{\sqrt{2\pi \rho}}
\nabla \cdot e
= \frac{\kappa}{4 \pi R^2} - \sum_{p}w_p \delta^2 (x- X_p)\,,
\qquad
e^i=f^{ij}(\dot{X}_p)_j.
\end{align}
The first equation is just Gauss's law on the sphere; to this order, the magnetic field perturbation $b$ is unsourced. The second equation in \eqref{eq_LorentzF} implies that vortices move with small drift velocities $|\vec{\dot{X}}_p|\sim 1/\sqrt{\rho}$ in trajectories of vanishing Lorentz force. This is consistent with the absence of \emph{fast} cyclotron degrees of freedom in the EFT. Integrating the first equation in \eqref{eq_LorentzF} we obtain
\begin{equation}
\sum_p w_p = \kappa\, .
\end{equation}
As expected, consistency of Gauss's law on the sphere implies the net charge of the vortices must neutralize the homogeneous contribution from the Euler term. Finally, it is easy to solve equation \eqref{eq_LorentzF} -- for example, for a single vortex ($\kappa=1$) placed on 
the north pole, one finds 
\begin{equation}\label{eq_singlevortexsol}
\alpha \frac{3/2}{\sqrt{2\pi \rho}}
e_{\rm sol}(\theta)
	=- \frac{1+\cos \theta}{4\pi R \sin \theta}\hat \theta\, .
\end{equation}
Solutions for $\kappa>1$ can be obtained by superposing \eqref{eq_singlevortexsol} with coordinates rotated to the location of the vortices. 

To leading order, the stress tensor is not affected by the presence of vortices and is still given by Eq.~\eqref{eq_T_original}. We then find the energy of the classical state as 
\begin{equation}
\Delta/R=\alpha \int d^2x\sqrt{-g}\left[(2\pi\rho)^{3/2}+
\frac{3/4}{\sqrt{2\pi \rho}}e^2\right]\,,
\end{equation}
with the electric $e$ field given by the solution to Gauss's law \eqref{eq_LorentzF}. Solving in terms of the vortex worldlines, we obtain:
\begin{equation}\label{eq_Delta_classical}
\Delta=
\frac{2}{3\sqrt{2\pi \chi_0}}Q^{3/2}+
\frac{\sqrt{2\pi \chi_0}}{8}\sqrt{Q}\left[
\kappa^2(\log 4-1)-\sum_{p, r}w_p w_r\log(\vec{R}_p-\vec{R}_r)^2\right]\,,
\end{equation}
where $\vec{R}_p=\left(\sin\theta_p\cos\phi_p,\sin\theta_p\sin\phi_p,
\cos\theta_p\right)$ is the position of the $p$th vortex in the $\mathbb{R}^3$ embedding of the unit sphere, and where the dimensionless susceptibility is $\chi_0 = \frac{1}{9\pi^3 \alpha^2}$ \eqref{eq_susceptibility}. The first term in parenthesis is the electrostatic energy due to the homogeneous charge induced by the Euler current, while the second one is the electrostatic potential for the interacting vortices. The terms with $p=p'$ diverge, but they will be cutoff at distances of order the vortex size $\delta X \sim \alpha^{-1}/\sqrt{\rho}$.\footnote{This estimate is obtained considering the distance from the vortex core at which the electric field \eqref{eq_singlevortexsol} becomes comparable with the leading monopole magnetic field.} These correspond to the usual fugacities of the vortices, which give a constant contribution $\sim \sqrt{\rho}\log \rho$ to the energy. Notice that because of the self-energy term, vortices with charge $w_p=\pm 1$ are generically energetically favored. We neglected contributions from higher derivative terms and from the particle masses in Eq.~\eqref{eq_Delta_classical}.  

Finally we can compute the angular momentum by evaluating the stress tensor \eqref{eq_T_original} on the solution \eqref{eq_singlevortexsol}. To leading order this is given by
\begin{equation}\label{eq_vortexJ}
\vec{J}=\frac{3\alpha}{2}\sqrt{2\pi\rho}\int d^2x\sqrt{g}\,\vec{n}^i\epsilon_{ij}\sqrt{g}e^j=\frac{Q}{2}\sum_p w_p \vec{R}_p\,,
\end{equation}
where we denoted collectively $\vec{n}^i=(n_x^i,n_y^i,n_z^i)$ the Killing vectors associated with rotations and we used Gauss's law to obtain the right hand side. When all vortices have the same charge, Eq.~\eqref{eq_vortexJ} implies that the angular momentum is proportional to the center-of-mass of the system. This property will be important in the following.

Finally we mention that in the EFT \eqref{eq_Stot} we neglected the possible existence of additional degrees of freedom living in the vortex cores. In particular, vortices may carry internal spin degrees of freedom. These may be included in the EFT straightforwardly introducing Grassmanian fields on the wordlines
% \cite{spin1,spin2,spin3}
and would provide additional $O(1)$ contributions to the angular momentum, as well as a $O(\sqrt{Q})$ contribution to the energy due to Pauli interaction between the spin and the monopole magnetic field (see \cite{Cuomo:2019ejv} for a more detailed analysis in a similar context). Both of these contributions are subleading with respect to the ones we consider in this paper, and therefore, in what follows, we shall consistently neglect any additional dynamics of the vortices other than the motion of the worldlines. We will nonetheless return to this point when considering microscopic realizations of the superfluid vortices for specific theories in Sec.~\ref{sec_appli}.

%######################################################################%
%======================================================================%
%======================================================================%
%======================================================================%
%######################################################################%
\section{Large charge CFT spectrum and OPEs}\label{sec_CFTdata}

In this section we apply the EFT discussed in the previous section to the study of the spectrum of conformal field theories, following the strategy presented in Ref.~\cite{Hellerman:2015nra}. More precisely, we consider generic non-parity preserving CFTs with a $U(1)$ global symmetry which enter a superfluid phase when coupled to a chemical potential $\mu \gg 1/R$ on $\mathbb{R}\times S^2$, where $R$ is the sphere radius. Such CFTs have a large charge spectrum of operators controlled by a superfluid EFT description. The EFT, combined with the state-operator correspondence, then allows to extract the CFT data of the corresponding operators in a systematic expansion in inverse powers of the charge.

\subsection{The lightest operator at fixed charge: classical analysis}

To leading order in $1/Q$, the dimension of the lightest operator of charge $Q$ is independent of the presence of vortices or higher derivative terms. It can be found simply from Eq.~\eqref{eq_T_original} and reads:
\begin{equation}\label{eq_DeltaQ}
\Delta_{\rm min}(Q)
 %\simeq \alpha \left(\frac{(2\pi)^{d+1}}{S_d}\right)^{1/d} Q^{\frac{d+1}{d}}\, .
 %\simeq \alpha \sqrt{2}\pi Q^{3/2}\, .
 \simeq \frac{2}{3\sqrt{2\pi \chi_0}} \, Q^{3/2} + \cdots\, ,
\end{equation}
where $\chi_0$ is the dimensionless charge susceptibility of the CFT. The notion of a charge susceptibility is meaningful regardless of the phase of matter that the CFT enters at finite density.%
	\footnote{For example the charge susceptibility of the $O(2)$ Wilson-Fisher CFT was obtained from Monte-Carlo simulations in Ref.~\cite{Banerjee:2017fcx} to be $\chi_0 = (\frac{2}{3c_{3/2}})^2 \approx 0.6225$.}
It is a property of the CFT that cannot be revealed by light operators only (it is similar in that respect to the coefficient $b_T$ in $d>2$ which controls the Cardy density of states of heavy operators \cite{Lashkari:2016vgj,Iliesiu:2018fao,Delacretaz:2020nit}). For CFTs that enter a superfluid phase at finite density, it is related to the superfluid stiffness by Eq.~\eqref{eq_susceptibility}. The result Eq.~\eqref{eq_DeltaQ} is independent of the discrete symmetries of the theory, and in fact follows from simple dimensional analysis \cite{Hellerman:2015nra}.\footnote{More precisely, Eq.~\eqref{eq_DeltaQ} follows from the assumption that the scaling dimension $\Delta_{\rm min}(Q)$ admits a non-trivial macroscopic limit \cite{Jafferis:2017zna}; this assumption might be violated when additional symmetries require the existence of flat directions, as in free theories or theories with extended supersymmetry \cite{Hellerman:2017veg}. } However, subleading corrections to \eqref{eq_DeltaQ},  the angular momentum of the ground state, and the spectrum of nearby operators are drastically affected by the parity-violating terms in the second line of \eqref{eq_Sfinal}, as we now discuss.

Let us now consider the classical ground state for the theory including vortices. This amounts to finding the classical configuration minimizing the expression \eqref{eq_Delta_classical}. We recall first that, because of their logarithmic self-energy, vortices with $|w_p|=1$ are energetically favored. Therefore we consider $\kappa$ identical charges interacting through the electrostatic field. The problem of finding the configuration of charges minimizing the logarithmic potential \eqref{eq_Delta_classical} on the sphere, or equivalently the product of their distances, is known as Whyte's problem \cite{dragnev2002discrete} in the mathematical literature. The solution is known for a few values of $\kappa$, but not in general\footnote{\label{footnote_kappas}Here are some examples: for $\kappa=2$ the charges are at opposite poles; for $\kappa=3$ they are on the corners of an equilateral triangle; for $\kappa=4$ they are on the corners of a tetrahedron.}. However, we shall need only the following general property of the solution \cite{Bergersen_1994}: the minimal energy configuration for $\kappa>1$ has center of mass at the origin of the sphere, i.e.~vanishing dipole. In terms of the superfluid and its vortices, this implies that the ground state has vanishing angular momentum \eqref{eq_vortexJ} for $\kappa>1$. For $\kappa=1$ instead, Eq.~\eqref{eq_vortexJ} implies that the ground state has spin $J=Q/2$. In summary the classical spin of the lightest operator of charge $Q$ reads:
\begin{equation}\label{eq_vortexJgroundstate}
J_{}(Q)=\begin{cases}
Q/2 & |\kappa|=1\,,\\
0 & |\kappa|\neq 1\,.
\end{cases}
\end{equation}
The leading correction to its dimension $\Delta_{\rm min}(Q)$ \eqref{eq_DeltaQ} comes from the fugacities of the vortices, which can be obtained from \eqref{eq_Delta_classical} without solving the interactions:
\begin{equation}\label{eq_DeltaQ_vort}
%\Delta_Q = 
	%\frac{2}{3\sqrt{2\pi \chi_0}} \, Q^{3/2} + \frac{1}{12 \sqrt{2} \pi \alpha }\sqrt{Q}\log Q + O(\sqrt{Q})\, .
\Delta_{\rm min}(Q) = 
	\frac{2}{3\sqrt{2\pi \chi_0}} \, Q^{3/2} + \kappa \frac{\sqrt{2\pi \chi_0}}{8} \sqrt{Q}\log \frac{Q}{\chi_0} + O(\sqrt{Q})\, .
\end{equation}
In words, each vortex gives a $\sqrt{Q}\log Q$ contribution to the dimension of the large charge operator. This is more singular than the non-universal $O(\sqrt{Q})$ corrections, which arise from the vortex tensions and interactions, as well as from the subleading terms in the EFT \eqref{eq_chiEFT}. {We do not write the $O(\sqrt{Q})$ term explicitly; its coefficient depends on additional Wilsonian coefficients of the EFT}. We stress that unlike in Ref.~\cite{Cuomo:2017vzg}, where it was found that the lightest state with charge $Q$ and spin $\sqrt{Q}\lesssim J < Q$ also contains vortices (with again a  $\sqrt{Q}\log Q$ contribution to $\Delta$), we are here discussing the lightest operator {at any spin} in a parity violating CFT. When $|\kappa|\neq1$, this state has no angular momentum but still contains vortices. Higher spin operators with the same charge $Q$ will also exist but have larger dimension -- they correspond to phonon or vortex excitations on top of the ground state and will be discussed in Sec.~\ref{ssec_spectrum_vortices}. We have kept track of the charge susceptibility inside the logarithm in Eq.~\eqref{eq_DeltaQ_vort}; although one expects $\chi_0=O(1)$ for strongly coupled CFTs, in weakly coupled theories $\chi_0$ can take parametrically large or small values as we will see in Sec.~\ref{sec_appli} (see also \cite{Badel:2019oxl}).

The present classical analysis leaves some open questions. For instance, though the ground state for $|\kappa|\neq 1$ has angular momentum $J=0$, it is classically inhomogeneous -- what is then its quantum mechanical spin? What is its degeneracy? 
And are there quantum corrections to Eqs.~\eqref{eq_vortexJgroundstate} and \eqref{eq_DeltaQ_vort}? Addressing these questions requires a quantum-mechanical description of the vortex worldlines. We shall illustrate how to proceed in Sec.~\ref{ssec_spectrum_vortices}, where we show that quantum corrections provide only subleading corrections to our results \eqref{eq_vortexJgroundstate} and \eqref{eq_DeltaQ_vort}, and discuss the low-lying spectrum of vortex excitations. We study in particular the dimension $\Delta_{\rm min}(Q,J)$ of the lightest operator at large charge $Q$ and finite spin $J$ -- lightest states at fixed quantum numbers are more easily accessible via numerical methods%
	\footnote{For example, it should be straightforward to extend the Monte-Carlo simulations of Ref.~\cite{Banerjee:2017fcx} to obtain $\Delta_{\rm min}(Q,J)$ for the first few spins. Although their current results are consistent with any phase of matter at finite charge and energy density, observing the (close to) linear dispersion $\Delta_{\rm min}(Q,J) = \Delta_{\rm min}(Q) + \sqrt{\frac{J(J+1)}{2}}$ would essentially confirm superfluidity. On a cubic lattice, spins $J=0,1,2,3,4$ should be accessible, since the cubic subgroup of $SO(3)$ has 5 irreps. In parity-violating CFTs with $\kappa\neq0$, we find that $\Delta_{\rm min}(Q,J)$ is instead very different and not controlled by phonons, see e.g.~Eq.~\eqref{eq_DeltaQJ_kappa2}.}.

Before doing so, we briefly review the spectrum of superfluid phonons, which arises even when $\kappa=0$ and whose treatment is almost identical to the parity-preserving case \cite{Hellerman:2015nra}.

%======================================================================%
%======================================================================%
%======================================================================%
\subsection{Superfluid phonons}\label{ssec_phononspectrum}

The dynamics of the superfluid phonons can be described expanding Eq.~\eqref{eq_EFT_intro} to quadratic order in the gauge field fluctuations:
\begin{equation}\label{eq_EFTpert_re}
\begin{split}
S
	&\simeq \alpha   \frac{3/4}{\sqrt{2\pi \rho}} \int d^3 x \left[e^2 - \frac12 b^2\right] + O(b^3/\rho) \\
	& + \frac{\kappa}{8\pi} \frac{1}{2\pi \rho} \int d^3 x \, \epsilon_{ij} e^i \dot e^j + O(\d b^3 / \rho)
	 + \cdots \, .
\end{split}
\end{equation}
If coefficients $\alpha,\, \kappa$ are order unity, both derivatives and interactions are suppressed by the same scale $\Lambda \sim \sqrt{\rho}$. The theory could be quantized as such, with $\kappa$ giving a contribution to the Goldstone propagator. However we prefer to remove the $\kappa$ term with a field redefinition; the quadratic action will then match that of parity-preserving superfluids so that the quantization of Refs.~\cite{Hellerman:2015nra,Monin:2016jmo} applies, at the cost of introducing new terms in operators such as the currents $j_\mu$ and $T_{\mu\nu}$. The appropriate field redefinition is $a_\mu \to a_\mu + \delta a_\mu$ with
\begin{equation}\label{eq_a_redef_pert}
\delta a_0 = \frac{\kappa}{8\pi \alpha}\frac{1/3}{\sqrt{2\pi\rho}} b \, , \qquad
\delta a_i = \frac{\kappa}{8\pi \alpha}\frac{2/3}{\sqrt{2\pi\rho}} \epsilon_{ij} e_j \, .
\end{equation}
Note that for the purposes of doing perturbation theory, the normalization $\oint \frac{a}{2\pi} \in \mathbb Z$ need not be preserved. The action is now given by only the $\alpha$ term in 
%\eqref{eq_EFTpert}
\eqref{eq_EFTpert_re}, i.e.~it is identical to that of parity-preserving conformal superfluids, which were quantized on the sphere in \cite{Hellerman:2015nra,Monin:2016jmo}. The canonically normalized Goldstone degree of freedom is mode-expanded on the sphere of radius $R$ as
\begin{equation}\label{eq_mode}
\pi_c(t,\hat n) = \pi_0 + \pi_1 t + \sum_{J>0}\sum_{|m|\leq J} \frac{1}{\sqrt{2\Omega_J}} \left(a_{Jm} Y_{Jm}(\hat n) e^{-i\Omega_J t/R} + \hbox{h.c.}\right)\, , 
\end{equation}
with creation and annihilation operators satisfying $[a_{Jm},a_{J'm'}^\dagger] = \delta_{JJ'}\delta_{mm'}$, and $\Omega_J = \sqrt{\frac{J(J+1)}2}$. The Hilbert space of the EFT in the sector of charge $Q = 4\pi R^2 \rho$ consists of phonon Fock states
\begin{equation}\label{eq_Fock}
|Q,\{n_{Jm}\}\rangle 
	= \prod_{J>0}\prod_{|m|\leq J} \frac{1}{\sqrt{n_{Jm}!}}(a_{Jm}^\dagger)^{n_{Jm}}|Q\rangle\, , 
\end{equation}
with energy $\frac{1}{R}\sum_{J\geq |m|}n_{Jm}\Omega_J$ above the ground state $|Q\rangle$. States with $n_{1m}> 0$ are descendants, and their matrix elements are determined in terms of the primaries from which they descend: in the EFT this follows from the momentum operator obtained from the linearized stress tensor %\eqref{eq_T_expand}
which satisfies $P_\mu \propto a^\dagger_{1m} + \hbox{h.c.}$. The spectrum receives corrections due to interactions; however at large $Q$ these are suppressed by $1/Q$. The fields $b,\,e_i$ appearing in \eqref{eq_EFTpert_re} are related to the canonically normalized scalar as
\begin{equation}\label{eq_eb_to_pic}
e_i = \left(\frac{\sqrt{2\pi\rho}}{3\alpha}\right)^{1/2} \epsilon_{ij}\d_j \pi_c + \cdots\, , 
	\qquad\quad 
b = -2 \left(\frac{\sqrt{2\pi\rho}}{3\alpha}\right)^{1/2} \dot \pi_c + \cdots \, .
\end{equation}
%

%======================================================================%
%======================================================================%
%======================================================================%
\subsection{Quantization and the spectrum of vortices}\label{ssec_spectrum_vortices}

We now turn to the vortices.
At a quantum level, the spectrum of vortices consists of Landau levels on the sphere, or monopole harmonics \cite{Wu:1976ge}. The lowest Landau level has spin
\begin{equation}
J 	= \frac12 \int_{S^2} \frac{f}{2\pi}
	= \frac{Q}{2}\, ,
\end{equation}
in agreement with the classical angular momentum obtained in Eq.~\eqref{eq_vortexJ}. Higher Landau levels are separated by a gap given by the cyclotron frequency $\omega_c \sim \rho / m_{\rm vortex} \sim \sqrt{\rho}/\gamma$ which is at the cutoff. This implies that the EFT describes $\kappa$ particles in the lowest Landau level. Quantization then imposes the following algebra for the vortex coordinates\footnote{Specifically, this follows from the vortex kinetic term $a_i \dot X^i_p$ in \eqref{eq_Sfinal}, see e.g. \cite{Cuomo:2017vzg}.}
\begin{equation}\label{eq_vortex_algebra}
[J^p_a, J^{p'}_b]
	= i \epsilon_{abc} J^{p}_c\delta_{pp'}\, , \qquad \quad \vec J^p =  \frac{Q}{2} \vec R^p\, .
\end{equation}
Because of the non-commutativity of the vortex coordinates, this system is sometimes referred to as a ``fuzzy sphere'' \cite{Hasebe:2010vp}. In the following, we will use Eq.~\eqref{eq_vortex_algebra} to study the spectrum of vortex excitations at a quantum level.

We start with the case $|\kappa|=1$, which is special in several aspects. Since there is only a single vortex, Eq.~\eqref{eq_Delta_classical} only produces the fugacity contribution accounted for in \eqref{eq_DeltaQ_vort}. The spectrum simply consists of a single spin $J=Q/2$ state, tensored with the spectrum of phonons. The most striking aspect of this result is that it implies that the lightest operator of charge $Q\gg 1$ has a large spin $J=Q/2${,  and therefore transforms non-trivially under rotations. Classically, the vortex could take any position on the sphere. Quantum mechanically, the large degeneracy of the ground state is resolved to be $\frac{Q}2(\frac{Q}2+1)$}. It is also interesting that this spin is half-integer for half the values of $Q$ -- a CFT that enters a superfluid phase with $|\kappa|=1$ is therefore necessarily fermionic. Notice that whether the ground state is fermionic for even or odd values of $Q$ depends on the internal spin of the vortex, and thus might depend on the specific theory under consideration. More generally, CFTs with $\kappa$ odd are necessarily fermionic.

Let us now turn to the case $|\kappa|>1$. The $\kappa$ vortices each have spin $J = Q/2$ and interact through the potential in \eqref{eq_Delta_classical}. 
We must therefore solve the problem of $\kappa$ spins with the following all-to-all interaction:
\begin{equation}\label{eq_H_vortex}
H
	= -\frac{\sqrt{2\pi \rho}}{3\alpha} \sum_{p<p'} \frac{1}{2\pi} \log |\vec J_p - \vec J_{p'}|^2 + \hbox{const.}\, .
\end{equation}
Let us start by considering $\kappa = 2$ vortices. In this case the Hamiltonian can be diagonalized in terms of the Casimir
\begin{equation}
(\vec J_1 + \vec J_2)^2 = J (J+1)\, , \qquad\quad
\vec J_1^{\,2} = \vec J_2^{\,2} = \tfrac{Q}2 \left(\tfrac{Q}2 + 1\right)\, , 
\end{equation}
and is given by
\begin{equation}\label{eq_Hkappa2}
H =-\frac{\sqrt{2\pi\rho}}{6\alpha\pi} \log \left[Q(Q+2) - J(J+1) \right]+\text{const.}\,,
\end{equation}
so that the ground state is unique and has spin $J=0$. Recall that classically, the ground state for $\kappa=2$ consists of two vortices at opposite pole -- the total angular momentum vanishes but the state is inhomogeneous and highly degenerate, as it can be rotated into other configurations. Here we see that at the quantum level the degeneracy is lifted and the true ground state minimizing \eqref{eq_Hkappa2} instead has spin $J=0$. It has energy given by \eqref{eq_DeltaQ_vort} with now $\kappa=2$. The low-lying vortex excitations  are states with $J>0$ -- expanding \eqref{eq_Hkappa2} for $J\ll Q$ shows that they have energy
\begin{equation}\label{eq_DeltaQJ_kappa2}
\Delta_{\rm min}(Q,J)
	= \Delta_{\rm min}(Q) + \frac{\sqrt{2\pi \chi_0}}{4} \frac{J(J+1)}{Q^{3/2}} + \cdots\,.
\end{equation}
These excitations are much softer than the superfluid phonons discussed in the previous section, which have $\Delta_{Q,\,J} - \Delta_{\rm min}(Q) = \sqrt{J(J+1)/2}$. The lightest operator of charge $Q$ and spin $0<J\lesssim Q$ is therefore an excitation of the vortices. The full spectrum consists of a tensor product of the vortex spectrum and the phonon Fock states -- part of this low-lying spectrum is shown in Fig.~\ref{fig_kappa2}. Notice also that we expect the two vortices to be identical, hence their wave-function should be properly symmetrized or antisymmetrized depending on their internal spin. For instance, assuming bosonic vortices, this implies that only even values of $J$ are allowed in Eq. \eqref{eq_DeltaQJ_kappa2}.

We were not able to solve the Hamiltonian \eqref{eq_H_vortex} exactly for larger values $|\kappa|>2$. Exact diagonalization of the Hamiltonian \eqref{eq_H_vortex} for low values of $\kappa$ and $Q$ suggests that the ground state minimizes its spin -- i.e. it has $J=1/2$ when $\kappa\cdot Q$ is odd and vanishes otherwise. In general, the commutation relations \eqref{eq_vortex_algebra} imply that the non-commutativity of the vortex coordinates is proportional to $1/Q$, which hence controls quantum corrections. This implies that the classical angular momentum \eqref{eq_vortexJ} may receive at most $O(1)$ corrections due to quantum effects. This may be seen more formally as follows: since the Hamiltonian \eqref{eq_H_vortex} describes $\kappa$ interacting spins with spin $Q/2$ and we are interested in the $Q\gg 1$ limit, it can be studied using the coherent state approximation \cite{Lieb1973}. This leads to the following representation for the partition function of \eqref{eq_H_vortex}:\footnote{Here we rescaled away the constant in front of Eq.~\eqref{eq_H_vortex} for notational simplicity.}
\begin{equation}
\begin{split}
Z[\beta]
	&= \Tr e^{-\beta H}
	= \Tr \left[ e^{\beta\sum_{p<p'}\log (\vec J_p - \vec J_{p'})^2}\right] \\
	&= \int D\vec N_p \, \delta(\vec N_p^2 - 1)
		\exp \left[  \beta\sum_{p<p'}\log (\vec N_p -  \vec N_{p'})^2 - \frac{\beta}{Q}\delta E(\vec N_p,\vec N_{p'}) + O \left(\frac1{Q^2}\right)\right], 
\end{split}
\end{equation}
where we introduced an inverse temperature $\beta$ to be taken to infinity. The coherent state representation amounts to a replacement of the spins $\vec J_p \to \frac{Q}{2} \vec N_p + O(Q^0)$ where the $\vec N_p$ are continuous unit vectors. The subleading in $1/Q$ pieces in the map depend on the operator under consideration, see Ref.~\cite{Lieb1973} for examples. The energy and angular momentum of the ground state can be obtained evaluating the partition function in the saddle-point approximation and taking the $\beta\rightarrow\infty $ limit. To leading order, this yields the same answer as the classical solution to Whyte's problem: $\vec N_p = \vec N_p^{\rm Whyte}$. However, the $\delta E/Q$ correction implies that it is advantageous to turn on a slight deviation $\vec N_p = \vec N_p^{\rm Whyte} + \delta \vec N_p$, with $\delta \vec N_p \sim 1/Q$, to minimize the energy. This may in turn produce a small magnetization.  

The spectrum of vortex excitations becomes richer as $\kappa$ increases. Although we will not attempt to study it in detail for $|\kappa|>2$, we expect low-lying vortex excitations to still be softer than phonons as we found for $|\kappa|=2$. Indeed, the dimension of the lightest large-charge operator with a spin $1\ll J \ll {Q}$ can be estimated classically by again considering a small deviation $\delta \vec N_p$ from the Whyte solution. This will produce a total spin $J \leq \kappa\frac{Q}2 \max_p|\delta N_p|$, and an energy cost $\delta E \sim \kappa^2\frac{\sqrt{\rho}}{\alpha} \delta N^2$. This leads to the estimate
\begin{equation}
\Delta_{\rm min}(Q,J) - \Delta_{\rm min}(Q)
	\simeq a_\kappa \sqrt{2\pi\chi_0} \frac{J^2}{Q^{3/2}}\, ,
\end{equation}
with $a_\kappa=O(1)$ for general $\kappa$ satisfying $1<|\kappa| = O(1)$.%
	\footnote{The $O(1)$ coefficient can be obtained from the explicit solution to Whyte's problem, when known. The $\kappa=2$ solution in Eq.~\eqref{eq_DeltaQJ_kappa2} gives $a_2 = \frac14$. For the $\kappa=3,4$ configurations described in footnote \ref{footnote_kappas}, we find $a_3=\frac13$ and $a_4=\frac{9}{16}$.} The fact that the spectrum of vortex excitations is softer than phonons also holds in the $\kappa\to \infty$ limit studied in Sec.~\ref{ssec_Vortex_Lattice}.

%======================================================================%
%======================================================================%
%======================================================================%
\subsection{Classical OPE coefficients for \texorpdfstring{$\kappa=1$}{k=1}}

The EFT can also be used to study matrix elements of light local operators in between the superfluid ground state and/or its excitations. In the following we shall analyze the predictions for some of the OPE coefficients involving the $U(1)$ current or the stress tensor, mostly focusing on the leading parity violating effects. When doing so, it will be convenient to distinguish between semiclassical correlators, whose leading value can be obtained by simply considering the classical expectation values of specific operators, and quantum ones, which are instead controlled by the (small) quantum fluctuations of the fields. We study the former here and the latter in Sec.~\ref{ssec_OPEphonon}.

In this section we first demonstrate how to compute classical OPEs, focusing on matrix elements of the $U(1)$ current in between the $|\kappa|=1$ ground state with $J=J_z=Q/2$. These can be obtained considering the classical expectation values of these operators for a configuration with a single vortex at the north-pole ($\theta=0$). For instance, from the expression for the $U(1)$ current \eqref{eq_currents_a}, we obtain the following matrix elements:
\begin{equation}\label{eq_j_mat_el}
\langle j_0\rangle=-\frac{Q}{4\pi R^2}\,,\qquad
\langle j_\phi\rangle=-\frac{2}{3\alpha}\frac{\sqrt{2\pi\rho}}{8\pi^2 R}(1+\cos\theta)\,.
\end{equation}
Here we used $f^\mu\simeq 2\pi\rho\delta^\mu_0$ and Eq. \eqref{eq_singlevortexsol}.

The predictive power of the EFT is easily illustrated comparing Eq.~\eqref{eq_j_mat_el} with the most general possible structure for the CFT matrix elements:\footnote{Eq. \eqref{eq_j_mat_el_CFT} can be obtained imposing rotational invariance on the sphere; the conformal group additionally relates these matrix elements to different ones involving descendant states, but it does not constraint further the expressions \eqref{eq_j_mat_el_CFT}.}
\begin{equation}\label{eq_j_mat_el_CFT}
\begin{split}
\langle \text{vor}_J | j_0 (x) | \text{vor}_J \rangle &= \begin{cases}
\sum_{m=0}^{J} a_m\sin^{2m}\theta
+\cos\theta\sum_{m=0}^{J-1} b_m\sin^{2m}\theta\,,\quad &J=\text{integer}\,,\\
\sum_{m=0}^{J-1/2} a_m\sin^{2m}\theta
+\cos\theta\sum_{m=0}^{J-1/2} b_m\sin^{2m}\theta\,,&J=\text{half-integer}\,,
\end{cases} \\[3pt]
\langle \text{vor}_J | j_\phi (x) | \text{vor}_J \rangle &= 
\begin{cases}
\sum_{m=1}^J c_m\sin^{2m}\theta
+\cos\theta\sum_{m=1}^{J} d_m\sin^{2m}\theta\,,
\quad &J=\text{integer}\,,\\
\sum_{m=1}^{J+1/2} c_m\sin^{2m}\theta
+\cos\theta\sum_{m=1}^{J-1/2} d_m\sin^{2m}\theta\,,&J=\text{half-integer}\,,
\end{cases}
\end{split}
\end{equation}
where $| \text{vor}_J\rangle$ is the $J=J_z$ ground state.
The coefficients $a_m$ and $c_m$ multiply parity-even structures, while the coefficients $b_m$ and $d_m$ parity odd ones. To compare this Eq.~with the EFT result \eqref{eq_j_mat_el}, we notice that not all the structures in Eq.~\eqref{eq_j_mat_el_CFT} can be predicted, as some of them will be peaked at the vortex core or they might involve Fourier components with frequency larger than the EFT cutoff $\sim\sqrt{\rho}$. It is in turn convenient to use half-angle formulas to rewrite the matrix elements in Eq.~\eqref{eq_j_mat_el_CFT} as:
\begin{equation}\label{eq_j_mat_el_CFT2}
\begin{split}
\langle \text{vor}_J | j_0 (x) | \text{vor}_J \rangle &=\sum_{m=0}^{\sim\sqrt{Q}} \alpha_m\cos^{2m}\frac{\theta}{2}+\text{terms outside the EFT}\,,\\
\langle \text{vor}_J | j_\phi (x) | \text{vor}_J \rangle &=\sum_{m=1}^{\sim\sqrt{Q}} \beta_m\cos^{2m}\frac{\theta}{2}+\text{terms outside the EFT}\,,
\end{split}
\end{equation}
The $\alpha_m$ and $\beta_m$ in Eq.~\eqref{eq_j_mat_el_CFT2} are linear combinations of the coefficients in Eq.~\eqref{eq_j_mat_el_CFT}.
Terms with $m\gtrsim \sqrt{Q}$ contain Fourier components with frequency larger than the cutoff and are exponentially suppressed away from the vortex core, therefore we neglected them in Eq.~\eqref{eq_j_mat_el_CFT2}. We may now compare with Eq.~\eqref{eq_j_mat_el} to obtain:
\begin{align}
\alpha_m=-\frac{Q}{4\pi R^2}\delta_m^0\,,\qquad
\beta_m=-\frac{\sqrt{\chi_0\,Q}}{2 \sqrt{2 \pi }R^2}\delta_m^1\,,\qquad\text{for}\qquad m\ll\sqrt{Q}\,.
\end{align}

We may similarly compute the matrix elements for the stress-energy tensor. Using Eq.~\eqref{eq_T_original} the EFT provides
\begin{equation}
\label{eq_T_mat_el}
\langle T_{00}\rangle\simeq\alpha(2\pi\rho)^{3/2}\,,\qquad
\langle T_{0\phi}\rangle\simeq\frac{2\pi\rho(1+\cos\theta)}{4\pi R}
\,,\qquad
\langle T_{\phi\phi}\rangle\simeq\frac{\alpha}{2}(2\pi\rho)^{3/2}\,,
\end{equation}
where we neglected subleading orders.
Proceeding as before, we compare Eq.~\eqref{eq_T_mat_el} with the following general parametrization:
\begin{equation}\label{eq_T_mat_el_CFT2}
\begin{split}
\langle \text{vor}_J | T_{00} (x) | \text{vor}_J \rangle &=\sum_{m=0}^{\sim\sqrt{Q}} \bar{\alpha}_m\cos^{2m}\frac{\theta}{2}+\text{terms outside the EFT}\,,\\
\langle \text{vor}_J | T_{0\phi} (x) | \text{vor}_J \rangle &=\sum_{m=1}^{\sim\sqrt{Q}} \bar{\beta}_m\cos^{2m}\frac{\theta}{2}+\text{terms outside the EFT}\,,\\
\langle \text{vor}_J | T_{\phi\phi} (x) | \text{vor}_J \rangle &=\sum_{m=0}^{\sim\sqrt{Q}} \bar{\gamma}_m\cos^{2m}\frac{\theta}{2}+\text{terms outside the EFT}\,.
\end{split}
\end{equation}
As a result we obtain:
\begin{equation}
\bar{\alpha}_m=2\bar{\gamma}_m=\frac{2  \,Q^{3/2}}{3\sqrt{2\pi \chi_0}R^2} \delta_m^0\,,\qquad
\bar{\beta}_m=\frac{Q}{4\pi R^2}\delta_m^1\,,\qquad
\text{for}\qquad m\ll\sqrt{Q}\,.
\end{equation}

Finally we comment that a purely classical approach cannot be used to compute OPE coefficients in the ground state for $\kappa= 2$. This is because it does not have macroscopic spin; in fact it has $J=0$, and there is no semiclassical approximation for the homogeneous quantum wave-function of the vortices. Instead, one must properly integrate over the zero-modes of the saddle-point configuration in the corresponding path-integral.\footnote{A similar procedure involving the zero mode of the Goldstone field ensures charge conservation in correlation functions \cite{Monin:2016jmo}.} A semiclassical approach does however allow to compute the OPE coefficients of the excited states \eqref{eq_DeltaQJ_kappa2} with $J\gg 1$ for $\kappa=2$. The procedure is analogous to the one described above and we do not report the results here. 

%======================================================================%
%======================================================================%
%======================================================================%
\subsection{Quantum OPE coefficients and transport}\label{ssec_OPEphonon}

\begin{figure}%[h]
\vspace{5pt}
\centerline{
\begin{overpic}[width=0.6\textwidth,tics=10]{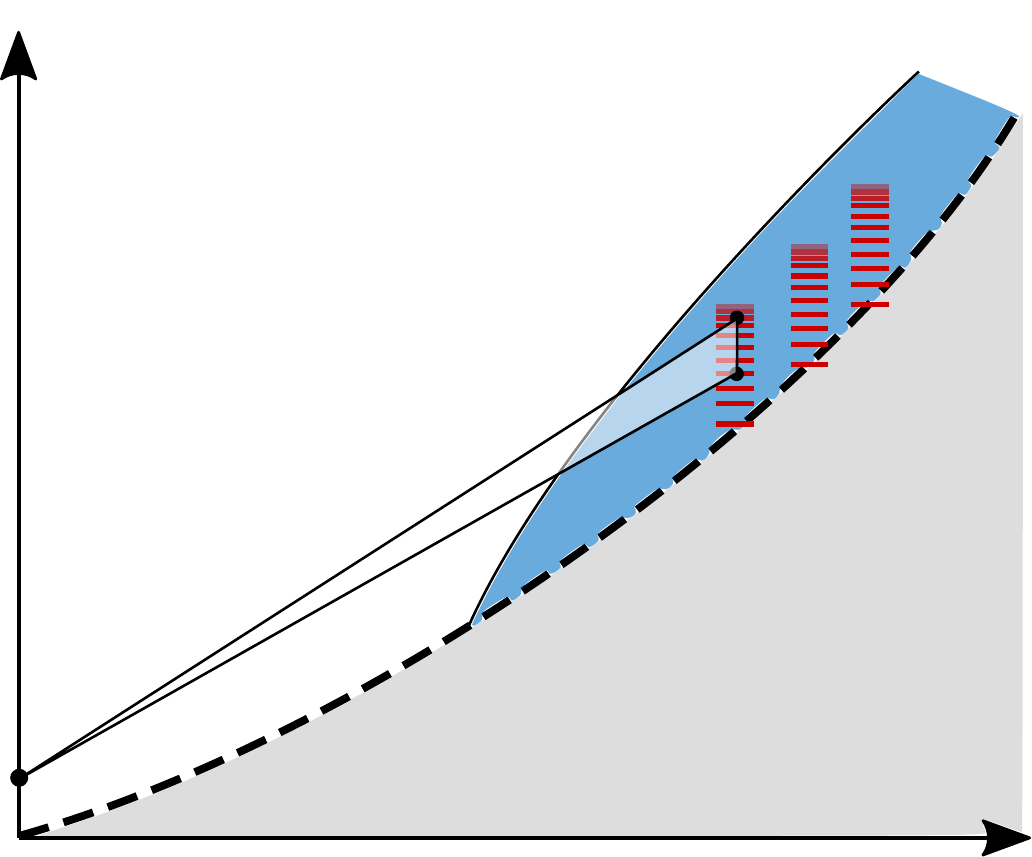}
	 \put (-1,85) {\Large$\Delta$} 
	 \put (-15,7) {\large$j_\mu,\,T_{\mu\nu}$} 
	 \put (105,-1) {\Large$Q$} 
	 \put (100,75) {\large$\sim Q^{\frac{d+1}{d}}$} 
\end{overpic} }
\caption{\label{fig_spectrum}
Spectrum in the large charge sector of a CFT. The triangle shows an example of a heavy-heavy-light OPE coefficient that can be computed from the superfluid EFT.
}
\end{figure}
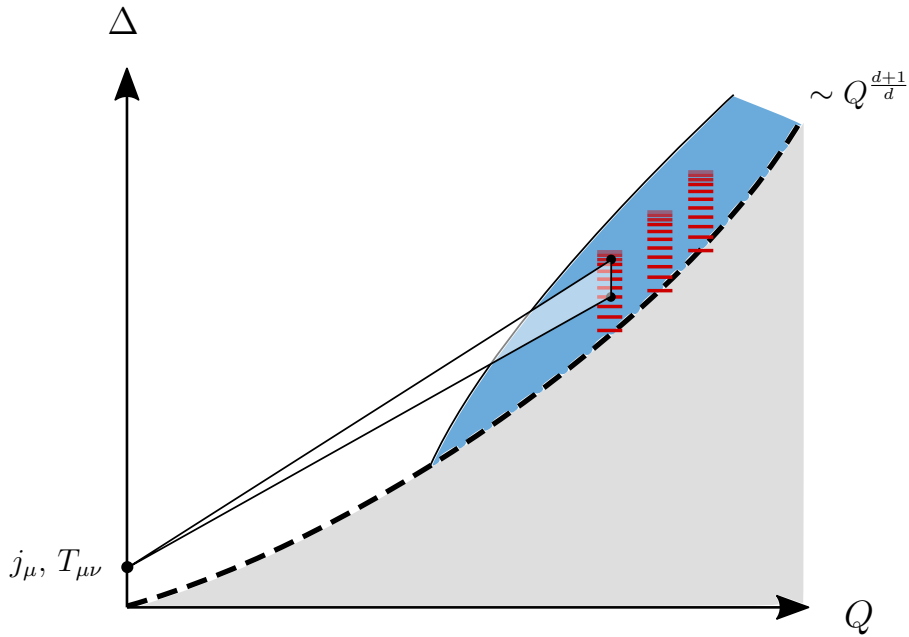

Phonon excitations lead to transport properties -- namely two-point functions of currents $j_\mu$, $T_{\mu\nu}$ in the finite density state -- that are characteristic of superfluids. One such feature for parity-violating superfluids is the Hall viscosity \eqref{eq_etatilde}, proportional to $\kappa$. In a CFT, transport can be studied with 4-point functions involving two heavy operators, e.g.~$\langle Q TT Q\rangle$. At a more basic level, transport signatures should be visible in off-diagonal heavy-heavy-light OPE coefficients $\langle Q T Q'\rangle$ \cite{Delacretaz:2020nit}. When the large charge sector is controlled by the superfluid EFT, we can make this manifest: the intermediate states in $\langle Q TT Q\rangle$ that give the dominant contribution to transport correspond to the ground state dressed with a single superfluid phonon -- choosing $Q'=Q_J\equiv a^\dagger_{Jm} |Q\rangle $ to be such a state (see Eq.~\eqref{eq_Fock}), one expects OPE coefficients $\langle Q j Q_J\rangle$ and $\langle Q T Q_J\rangle$ to capture the salient features of transport in the superfluid state. This is illustrated in Fig.~\ref{fig_spectrum}. Similar OPE coefficients were studied in Refs.~\cite{Monin:2016jmo,Jafferis:2017zna}; we follow the strategy presented there to compute heavy-heavy-light OPE coefficients from the EFT. We are focusing on privileged light operators -- the $U(1)$ current $j_\mu$ and stress tensor $T_{\mu\nu}$ -- which are universally present in the CFTs of interest. Moreover, their fixed normalization implies that matching these operators to the EFT only involves, to leading order, the charge susceptibility $\chi_0$ (or equivalently $\alpha$ \eqref{eq_susceptibility}) which can already be read off from the spectrum \eqref{eq_DeltaQ}.

We therefore proceed to computing $\langle Q j Q_J\rangle$ and $\langle Q T Q_J\rangle$, focusing in this section on phonon physics. At small wavevectors $k\sim 1/R$ on the sphere, phonons will be sensitive to the presence of the $\kappa$ vortices. To cleanly separate the phonon dynamics from that of vortices, which were studied in the previous sections, we focus on wavevectors $k\gg 1/R$, i.e~on operators $Q_J$ with spin 
\begin{equation}
1 \ll J \lesssim \sqrt{Q} \, .
\end{equation}
In this regime, we can ignore curvature corrections, e.g.~to the stress tensor \eqref{eq_T}. As discussed in Sec.~\ref{ssec_phononspectrum}, in order to utilize the phonon algebra of Refs.~\cite{Hellerman:2015nra,Monin:2016jmo,Jafferis:2017zna}, we needed to perform a field redefinition  \eqref{eq_a_redef_pert}. After this field redefinition the $U(1)$ current is no longer given by $j_\mu = \frac{1}{2\pi} \epsilon_{\mu\nu\lambda} \d^\nu a^\lambda$ but rather, to linear order in $b,\,e_i$, by
\begin{equation}\label{eq_j_expand}
j^0 = \rho - \frac{1}{2\pi} b \, , \qquad
j^i = -\frac{1}{2\pi} \epsilon^{ij}\left(e_j + \frac{2/3}{\sqrt{2\pi \rho}} \frac{\kappa}{8\pi \alpha}\d_j b\right)\, . 
\end{equation}
The stress tensor also receives an additional contribution shown in Appendix \ref{sapp_T_lin}. It is then straightforward to compute matrix elements of the current between single phonon states and the vacuum, by expressing $e_i$ and $b$ in terms of the canonical phonon field \eqref{eq_eb_to_pic} and using the algebra \eqref{eq_mode}. For the time component of the current, one obtains:
\begin{equation}\label{eq_j0_matrixel}
\begin{split}
\langle Q | j^0(t,\hat n) | Q,Jm\rangle
	&= -\frac{1}{2\pi} \langle Q | b \,a^\dagger_{Jm} | Q\rangle \\ 
	&\simeq -i\sqrt{\Omega_J} \frac{(\chi_0 \rho/2)^{1/4}}{R} Y_{Jm}(\hat n) e^{-i\Omega_J t /R} \, .
\end{split}
\end{equation}
where in the second step we used \eqref{eq_eb_to_pic} and \eqref{eq_mode}. For the spatial component,
\begin{equation}\label{eq_ji_matrixel}
\begin{split}
\langle Q | j^i(t,\hat n) | Q,Jm\rangle|_{\rm odd}
	&\simeq -\frac{1}{2\pi} \frac{2/3}{\sqrt{2\pi \rho}} \frac{\kappa}{8\pi \alpha}\langle Q | \d_j b \,a^\dagger_{Jm} | Q\rangle \\ 
	&\simeq -i\kappa\sqrt{\Omega_J} \frac{(2 \chi_0^3 /\rho)^{1/4}}{8R^2} \d_jY_{Jm}(\hat n) e^{-i\Omega_J t /R} \, ,
\end{split}
\end{equation}
where we focused on the parity-odd part, proportional to $\kappa$; the parity-even part can be computed similarly and is related to \eqref{eq_j0_matrixel} by the Ward identity. Matrix elements for the stress tensor are obtained in Appendix \ref{sapp_T_lin}.

These expressions can be recast into OPE coefficients after choosing a basis for CFT three-point functions. On general grounds we expect the three-point functions to involve one parity-even and one parity-odd OPE coefficient, see Ref.~\cite{Costa:2011mg}. In their notation, the three-point function of $\mathcal{O}_1 = Q^\dagger$, $\mathcal{O}_2 = j_\mu$ and $\mathcal{O}_3 = Q_J$ takes the form
\begin{equation}\label{eq_OPE_CFT}
\begin{split}
G(P_1,P_2,P_3; Z_1,Z_2,Z_3)
	&= \langle Q^\dagger \,j \, Q_{J}\rangle|_{\rm even} \sum_{\alpha=0}^{J_2} c_\alpha(J_2,J_3)\frac{V_2^{J_2 - \alpha} H_{23}^\alpha V_3^{J_3 - \alpha}}{P_{12}^{\tau_{12,3}/2}P_{23}^{\tau_{23,1}/2}P_{31}^{\tau_{31,2}/2}} \\
	&\ + \langle Q^\dagger \,j \, Q_{J}\rangle|_{\rm odd} \sum_{\alpha=0}^{J_2-1} \tilde c_\alpha(J_2,J_3)\frac{\epsilon_1 V_2^{J_2 - \alpha-1} H_{23}^\alpha V_3^{J_3 - \alpha-1} }{P_{12}^{\tau_{12,3}/2}P_{23}^{\tau_{23,1}/2}P_{31}^{\tau_{31,2}/2}} \, ,
\end{split}
\end{equation}
with $J_2 = 1$ and $J_3 = J$ (recall that $J>1$ for the single phonon state to be a primary). The three point function with $\mathcal{O}_2 = T_{\mu\nu}$ is similar, with $J_2=2$. The coefficients $c_\alpha$ and $\tilde c_\alpha$ are entirely fixed up to two overall constants $c_0$ and $\tilde c_0$ from the Ward identities. In order not to lose the reader to technical details, we will not specify the overall normalization conventions (which depend on the spin $J$) -- the OPEs listed below therefore have an overall $J$-dependent normalization which we are not keeping track of (their relative $J$ dependence is meaningful however). We are now left with the dynamics, in the OPE coefficients
\begin{equation}
\langle Q^\dagger \,j \, Q_{J}\rangle|_{\rm even} \, , \qquad
\langle Q^\dagger \,j \, Q_{J}\rangle|_{\rm odd} \, , \qquad
\langle Q^\dagger \,T \, Q_{J}\rangle|_{\rm even} \, , \qquad
\langle Q^\dagger \,T \, Q_{J}\rangle|_{\rm odd} \, .
\end{equation}
We only report the result to leading order in $Q\gg 1$ and for spins $\sqrt{Q} \gtrsim J \gg 1$, dropping numerical factors and emphasizing instead the parametric dependence on the quantum numbers $Q,\,J$ and CFT properties $\chi_0,\,\kappa$. The OPE coefficients involving the current $j_\mu$ can be obtained from Eqs.~\eqref{eq_j0_matrixel} and \eqref{eq_ji_matrixel}:
\begin{equation}\label{eq_j_QuantumOPE}
\begin{split}
\langle Q^\dagger \,j \, Q_{J}\rangle|_{\rm even}
	&\sim (\chi_0 Q)^{1/4} J^{1/2}\, , \\
\langle Q^\dagger \,j \, Q_{J}\rangle|_{\rm odd}
	&\sim \kappa \chi_0^{3/4} Q^{-1/4} J^{3/2} \, .
\end{split}
\end{equation}
For the stress tensor, one finds 
\begin{equation}\label{eq_T_QuantumOPE}
\begin{split}
\langle Q^\dagger \,T \, Q_{J}\rangle|_{\rm even}
	&\sim \frac{Q^{3/4}}{\chi_0^{1/4}} {J}^{1/2} \, , \\
	\langle Q^\dagger \,T \, Q_{J}\rangle|_{\rm odd}
	&\sim \kappa \chi_0^{1/4} Q^{1/4} J^{3/2} \, .
\end{split}  
\end{equation}
To leading order in $1/Q$, both parity-even OPE coefficients only depend on the quantum numbers of the operators ($Q$ and $J$) and on the susceptibility $\chi_0$ -- they are entirely fixed once $\chi_0$ is measured through \eqref{eq_DeltaQ}. The leading parity-odd OPE coefficients involve a new parameter: the quantized coefficient $\kappa\in \mathbb Z$ appearing in the EFT \eqref{eq_aEFT_theta}. Eqs.~\eqref{eq_j_QuantumOPE} and \eqref{eq_T_QuantumOPE} exemplify how transport can be captured by off-diagonal heavy-heavy-light OPE coefficients -- for example the product of the two lines in \eqref{eq_T_QuantumOPE} computes the Hall viscosity \eqref{eq_etatilde} of the superfluid.

%======================================================================%
%======================================================================%
%======================================================================%
\subsection{Large \texorpdfstring{$\kappa$}{k} and vortex lattices}\label{ssec_Vortex_Lattice}

So far, we implicitly considered $\kappa$ as an $O(1)$ parameter. It is natural to ask what happens in the limit $\kappa\gg 1$. Though we expect such situation to be mostly relevant in weakly coupled theories, such as the ones we discuss in sec. \ref{sec_appli}, here we provide some general comments from the EFT perspective.

Let us focus first on the properties of the ground state. The limit $\kappa\rightarrow\infty$ of Whyte's problem received considerable attention in the mathematical literature \cite{Bergersen_1994,Saff1997DistributingMP}. It is simple to understand the main feature of the solution in our setup. For $\kappa\gg 1$ the vortex distribution may be treated as approximately continuous. To minimize the energy, it is then convenient for the vortices to arrange homogeneously on the sphere:
\begin{equation}
\rho_v(x)\equiv\sum_p \delta^{2}(x^i-X^i_p)\stackrel{\kappa\gg1}{\approx}\frac{\kappa}{4\pi R^2}
\end{equation}
where the last equality holds in the sense of distributions clearly. In this way their contributions cancels the charge sourced by the Euler term in the Gauss's law \eqref{eq_LorentzF}, which thus gives $e^i\approx 0$, hence minimizing the electrostatic energy. It may be shown that the contribution to the electrostatic energy from pairwise interactions in this limit reads \cite{Bergersen_1994,Saff1997DistributingMP}:
\begin{equation}
-\sum_{p\neq r}
\log (\vec{R}_p-\vec{R}_r)^2=-\kappa^2(\log 4-1)-\kappa\log \kappa+O\left(\kappa\right)\,.
\end{equation}
The leading term in this formula cancels the electrostatic energy due to the homogeneous charge sourced by the Euler current in eq. \eqref{eq_Delta_classical}, as expected. The logarithmic $-\kappa\log\kappa$ correction then recasts our result \eqref{eq_DeltaQ_vort} for the ground state energy in the form:
\begin{equation}\label{eq_DeltaQ_vort_largeK}
\Delta_{\rm min}(Q) = 
	\frac{2}{3\sqrt{2\pi \chi_0}} \, Q^{3/2} + \kappa \frac{\sqrt{2\pi \chi_0}}{8} \sqrt{Q}\log \frac{Q}{\kappa\,\chi_0} + O(\sqrt{Q})\,.
\end{equation}
For the EFT to hold, we need to require that the one derivative terms in the second line of the action \eqref{eq_Sfinal} be subleading with respect to the leading order action; this leads to the condition $\kappa\ll \alpha\sqrt{Q}\sim\sqrt{Q/\chi_0}$.\footnote{This condition is generically stronger than the one which follows from the requirement that vortex cores are not overlapping: $\kappa\ll \chi_0^2 Q$ (recall the comments below Eq. \eqref{eq_Delta_classical} on the vortex size).}
In practice Eq. \eqref{eq_DeltaQ_vort_largeK} might be of interest in weakly coupled theories, where the Wilson coefficients also depend explicitly on $\kappa$. In that case, it might happen that the subleading contribution from the vortex masses and the higher derivative terms are numerically more important than the logarithmic fugacity of the vortices in eq. \eqref{eq_DeltaQ_vort_largeK} for non-exponentially large charge.. 

The large $\kappa$ limit provides us with a handle on the otherwise complicated spectrum of excited states. It was indeed observed numerically that the configuration minimizing the potential energy takes a regular structure in this limit \cite{Saff1997DistributingMP}, analogous to the usual triangular lattice which is observed in superfluids \cite{tkachenko1966vortex,PhysRevLett.43.214}. %\footnote{More precisely, the configuration looks precisely as a triangular lattice, whose dual is made of hexagonal cells, up to 12 cells which are instead pentagonal to account for the non trivial Euler characteristic of the sphere. This is analogous to the classical patter of a football (soccer-ball for American readers), which has twenty hexagonal faces and twelve pentagonal ones.}
Therefore, at distances much larger than the typical vortex separation $1/\sqrt{\rho_v}\sim R/\sqrt{\kappa}$, we may resort to a simplified description in terms of the collective lattice coordinates of the vortices to study the CFT spectrum.

Rather than constructing the generic vortex lattice action, we prefer to adapt the vortex lattice EFT of \cite{Moroz:2018noc} to our problem; we do not expect significant differences in the most general case. We therefore consider the following quadratic Lagrangian:\footnote{Here we neglected an higher derivative term $-\frac{\kappa}{8\pi (2\pi\rho)}e_i\epsilon^{ij}\dot{e}_j$ from the expansion of the Euler current as well as a $\sim \sqrt{2\pi\rho}\,\dot{u}^i\dot{u}_i$ contribution to the kinetic term of the lattice coordinate fluctuations.}
\begin{equation}\label{eq_vortex_lattice_action}
\begin{split}
\mathcal{L}/\sqrt{g}&\simeq 
 \alpha   \frac{3/4}{\sqrt{2\pi \rho}}\left(e_i^2-\frac{1}{2}b^2\right)
+\frac{\kappa}{4\pi R}e_iu^i-\frac{\kappa (2\pi\rho)}{8\pi}u^i\dot{u}^j\epsilon_{ij} \\
&-\frac{C_1}{2}\sqrt{2\pi\rho}(\nabla_i u^i)^2
-\frac{C_2}{2}\sqrt{2\pi\rho}\left[(\nabla_i u^k)^2-\mathcal{R}_{ij}u^i u^j\right]\,,
\end{split}
\end{equation}
where $u^i$ is a vector representing the displacement from equilibrium of the vortex lattice coordinates. 
The first term is just the phonon action, whose spectrum is unmodified to leading order. The terms proportional to $\kappa$ arise from the Euler term and from the minimal coupling of the vortices to the gauge field. The coefficients $C_i$'s parametrize the elasticity of the solid. We can estimate their value from the expression \eqref{eq_Delta_classical} for the electrostatic potential; this gives $C_i\sim\kappa/\alpha$. Finally notice the appearance of the curvature $\mathcal{R}_{ij}$ in the last term,\footnote{This arises when expanding solid invariants -- see \cite{Esposito:2017qpj} for details about the solid EFT on the sphere.} whose role will become clear in a moment. 

Imposing that the determinant of the inverse propagator obtained from eq. \eqref{eq_vortex_lattice_action} vanishes, one finds the phonon mode, as well as a new mode with the following dispersion relation:
\begin{equation}
\omega^2_J=\frac{2\,C_2 }{3  \alpha (2\pi\rho R^2) }\left[\frac{J(J+1)}{R^2}- \mathcal{R}\right]\left[1+O\left(\frac{1}{\kappa}\right)\right]\,,\qquad
J\geq 1\,.
\end{equation}
The requirement $J\geq 1$ arises since vector fields have no zero mode on the sphere. The curvature contribution $\mathcal{R}=2/R^2$ implies that $\omega_J$ vanishes for $J=1$. This was to be expected, since the lattice coordinates can be thought as the Goldstone bosons of the rotations broken by the lattice. This implies that, analogously to the descendants created by the $J=1$ mode of the phonon, the rotation generators are proportional to the $J=1$ mode of the lattice fluctuations, which should hence be gapless. Finally, we notice that the sound speed of this mode is much smaller than the one of the phonons. Therefore, as we found previously for $\kappa=2$, the lightest mode with non-zero angular momentum is provided by fluctuations of the vortices and its energy reads:
\begin{equation}\label{eq_Delta_vortex_lattice_excitations}
\Delta_{\rm min}(Q,J)=\Delta_{\rm min}(Q)+C\sqrt{\frac{\chi_0\kappa}{Q}}\sqrt{J(J+1)-2}\,,\qquad
1\leq J^2\ll\kappa\,,
\end{equation}
where $C$ is a $O(1)$ Wilson coefficient. The prediction \eqref{eq_Delta_vortex_lattice_excitations} holds as long as the angular momentum does not exceed the scale set by the vortex density.

A remark is in order. The reader familiar with spinning superfluids might find the spectrum we just discussed rather bizarre. In particular, it is known that the low energy spectrum of a vortex lattice usually consists of a single gapless mode, with quadratic dispersion relation: the Tkachenko mode \cite{tkachenko}. However, that result applies in a regime of frequency which is inaccessible on the sphere for us, namely $\omega\ll m_{\rm Kohn}$ where $m_{\rm Kohn}$ sets the gap of the so called Kohn mode \cite{PhysRev.123.1242}. In our notation $m_{\rm Kohn}\sim \frac{\rho_v}{\alpha\sqrt{\rho}}$, hence $m_{\rm Kohn}\ll 1/R$ in our setup (see the discussion below Eq. \eqref{eq_DeltaQ_vort_largeK}). The two modes we find may instead be considered as the extrapolation of the Tkachenko and the Kohn mode to higher frequencies $m_{\rm Kohn}\ll\omega\ll\rho_v$.

Finally we notice that a similar EFT may also describe excitations of the large spin states studied in \cite{Cuomo:2017vzg}, with appropriate modifications since the vortex density is not homogeneous in that situation\footnote{We thank Angelo Esposito for discussions on this topic.}.

%######################################################################%
%======================================================================%
%======================================================================%
%======================================================================%
%######################################################################%
\section{Applications}\label{sec_appli}

We expect our results to apply to a number of 3d CFTs, both strongly and weakly interacting. Candidates without parity symmetry include QED${}_3$ with an odd number $N_f$ of fermions, or Chern-Simons--matter theories. Below we study a theory that has another perturbative handle, where our predictions can be tested and extended to regimes beyond the EFT discussed in the previous section. A number of perturbative checks have been made for parity preserving theories, see e.g.~\cite{delaFuente:2018qwv,Alvarez-Gaume:2019biu,Badel:2019oxl}.
	
%======================================================================%
%======================================================================%
%======================================================================%
\subsection{Anyon superfluid}\label{ssec_anyon}

We consider a weakly coupled parity-violating CFT: a single Dirac fermion coupled to $U(1)_{-k+\frac12}$ Chern-Simons term%
	\footnote{The half-integer level refers to the fact that the theory is regularized such that gapping out the fermion with $+m \bar\psi \psi$ ($-m\bar\psi\psi$) drives the system to a topological phase $S=-\frac{k'}{4\pi}\int \,ada$ with level $k'=k-1$ ($k'=k$). Since we will work to leading order in $k\gg 1$, we will neglect this subtlety in what follows.}
\begin{equation}\label{eq_S_anyon}
S
	= \int d^3x\, \bar\psi i\cancel D \psi - \frac{k}{4\pi} \epsilon^{\mu\nu\lambda} a_\mu\d_\nu a_\lambda\, , 
\end{equation}
with $\cancel D = \gamma^\mu (\d_\mu - i a_\mu)$. When the level $k$ is large, $1/k$ acts as a loop counting parameter and the theory is weakly coupled. For $k=0$, this theory has time-reversal symmetry (which acts non-trivially on the fermion field, mapping it to a monopole operator) and has been conjectured to flow to the $U(1)$ Wilson-Fisher CFT \cite{PhysRevB.48.13749,Seiberg:2016gmd,Karch:2016sxi}. For generic values of $k$ it is however not time-reversal invariant. Moreover, non-relativistic versions of this theory are known to enter a superfluid at finite density, at least at large $k$ \cite{PhysRevLett.60.2677,Chen:1989xs,PhysRevLett.65.2070}. We will in fact show that in this limit the EFT \eqref{eq_EFT_intro} can be {\em derived} from the microscopic weakly coupled CFT \eqref{eq_S_anyon} at finite density, confirming the validity of the framework in this context and providing the EFT coefficients $\chi_0$ and $\kappa$.

%======================================================================%
\subsubsection*{Anyons on the plane}

The theory \eqref{eq_S_anyon} has a global $U(1)$ symmetry carried by the current
\begin{equation}
j^\mu = \frac{1}{2\pi} \epsilon^{\mu\nu\lambda} \d_\nu a_\lambda\, .
\end{equation}
The equation of motion for $a_0$ is a constraint equation that attaches flux to fermion density
\begin{equation}\label{eq_anyonconstraint}
\bar\psi \gamma^0 \psi
	= k j^0\, .
\end{equation}
We start by studying the theory on the plane $\mathbb R^3$, at finite density  $\langle j^0\rangle = \rho $.
Eq.~\eqref{eq_anyonconstraint} then implies that the theory consists of a finite density $n_{\psi}\equiv \langle\bar\psi \gamma^0 \psi\rangle = k \rho$ of fermions in a background magnetic field $b_0 = 2\pi \rho$. When $k\to \infty$, the fermions decouple from the fluctuating gauge field $a_\mu$ so that the finite density ground state consists of $k$ filled Landau levels. 
Since the fermions are now gapped, they can be integrated out to yield 
a local effective action for $a_\mu$. Specifically, the action will admit a natural derivative expansion for momenta much smaller than $\sqrt{\rho/k}$, which sets the gap of the lightest particle-hole excitations of the fermionic field. The Hall response of $k$ filled Landau levels is $\sigma_{xy} = \frac{k}{2\pi}$ -- this implies that the leading term coming from integrating out the fermions precisely cancels the CS term in \eqref{eq_S_anyon}%
	\footnote{The sign can be checked at the classical level from the Lorentz force: $\d_\mu T^{\mu i } = f^{i\nu}j_\nu^\psi = 0$ $\Rightarrow$ $j^\psi_i = \frac{k}{2\pi} \epsilon_{ij} f_{j0}$ leading to a term in the effective action $+\frac{k}{4\pi}\epsilon^{\mu\nu\lambda}a_\mu\d_\nu a_\lambda$.
	}, 
so that the effective action for $a_\mu$ starts at two-derivatives and the photon is massless. This is the superfluid Goldstone boson. Since this is a CFT which enters a superfluid phase at finite density, we expect it to be described by the EFT in Sec.~\ref{sec_EFT}, with $\kappa\neq0$ since the CFT is parity-violating. To derive the EFT, we need to integrate out the Landau level fermions. This was done in Ref.~\cite{PhysRevB.95.085151}, which obtained to leading order in fields and derivatives 
\begin{equation}\label{EFT_anyon_flat}
S_{\rm eff}
	= \alpha_k\frac{3/4}{\sqrt{2\pi \rho}}\int d^3 x \left(e^2 - \frac12 b^2\right) + \cdots \, 
\end{equation}
with $\alpha_k = \frac{\sqrt{2}}{3\pi}k^{3/2} + O(k^{1/2})$. This matches the leading term in the expansion of the EFT \eqref{eq_EFTpert_re} (in particular the speed of sound $c_s = 1/\sqrt{2}$, as expected for conformal superfluids), with susceptibility \eqref{eq_susceptibility} given by
\begin{equation}\label{eq_chianyon}
\chi_0 
	= \frac{1}{9\pi^3 \alpha_k^2}
	= \frac{1}{2\pi k^3} + O(1/k^4)\, .
\end{equation}
The fact that $\chi_0$ is small at large $k$ implies that the strong coupling scale of the EFT is parametrically larger than the scale set by the density
\begin{equation}
\Lambda_{\rm sc} \sim \frac{\sqrt{\rho}}{\chi_0^{1/6}} \sim \sqrt{\rho k} \gg \sqrt{\rho}\, .
\end{equation}
However, as already commented, we do expect to see `new physics' -- e.g.~in the form of roton states \cite{Du:2020gqf} -- at energies $\Lambda\sim\sqrt{\rho / k}$. The large charge spectrum of the CFT will therefore contain operators with dimension $\sim\Delta_{\rm min}(Q) + \sqrt{Q/k}$ corresponding to these gapped excitations on top of the superfluid ground state.

Extending the calculation of \cite{PhysRevB.95.085151} to higher point-functions and higher derivatives, one could determine all the coefficients of the EFT \eqref{eq_EFT_intro} in flat space. By Weyl invariance this would determine also the bulk action for the theory in $\mathbb{R}\times S^2$ as an expansion in $1/R\Lambda$, where $\Lambda\sim\sqrt{\rho/k}$ is the EFT cutoff. In particular, we could find the parity-odd coefficient $\kappa$ in this way. However this is tedious to do in practice; we will instead discuss a simpler and more direct method to match its value below, where we discuss the theory on the sphere.

%======================================================================%
\subsubsection*{Anyons on the sphere}

Let us now consider the theory \eqref{eq_S_anyon} on $\mathbb{R}\times S^2$. The previous discussion immediately allows us to obtain the leading order dimension of monopole operators of large charge $Q$ in the theory \eqref{eq_S_anyon}. Indeed using the value \eqref{eq_chianyon} for the susceptibility and \eqref{eq_DeltaQ}, we obtain:
\begin{equation}\label{eq_anyonDeltaQ}
\Delta_{\rm min}(Q) \simeq 
	\frac{2}{3}(kQ)^{3/2} + \cdots
\end{equation}
On general grounds, we expect that this equation will receive $1/k$ relative corrections from higher loops in perturbation theory, as well as $k/Q$ relative corrections from higher derivative terms in the Lagrangian. We shall see that to leading order in the coupling Eq.~\eqref{eq_anyonDeltaQ} is actually exact. To show this, let us directly compute the dimension of monopole operators in the theory \eqref{eq_S_anyon} to leading order in coupling $k$, but all orders in $Q/k$, i.e. in the double-scaling limit $k\rightarrow \infty$, $Q\rightarrow\infty$ with $Q/k$ fixed, in analogy with \cite{Badel:2019oxl,Grassi:2019txd}.\footnote{A similar double-scaling limit can be studied by a straightforward generalization of the ideas discussed in \cite{Badel:2019oxl} in Chern-simons theories coupled to bosonic matter; similar ideas were explored in~\cite{Watanabe:2019adh} in a $SU(2)$ Chern-Simons theory coupled to a scalar field.} As in those works, this is achieved by means of the state-operator correspondence. To leading order, the gauge field is set to a background value $\bar{a}_\mu$, which provides the homogeneous magnetic field on the sphere. The Dirac field in a monopole background is then quantized using spinor monopole harmonics (see e.g. \cite{Chester:2017vdh} for a definition). The constraint \eqref{eq_anyonconstraint} implies that the lowest energy state of charge $Q$ consists of $kQ$ particles organized in Landau levels. Summing their energies gives:
\begin{equation}\label{eq_Deltadouble}
\Delta_{\rm min}(Q)=\frac{2}{3}k^3\left(\frac{Q}{k}\right)^{3/2}\left[1+O\left(\frac{1}{k}\right)\right]\,.
\end{equation}
This equation coincides with \eqref{eq_anyonDeltaQ};  this result is exact in $Q/k$ to leading order in the coupling.\footnote{This result agrees with \cite{Radicevic:2015yla,Chester:2017vdh}, where the authors worked in the limit of $Q$ fixed.} Fluctuations of the gauge field couple to particle-hole excitations of this state, whose gap depends only on $Q/k$ up to $O(1/k)$ corrections. Since this gap does not scale with $k$, the weak coupling of the theory ensures that fluctuations are indeed suppressed. Of course, in principle one can compute all corrections systematically integrating out the Dirac field on a finite charge state as in \cite{PhysRevB.95.085151}. In practice, as we showed before, the gauge field propagator receives large corrections in this limit. These make it non-local at short distances, and the calculation is technically challenging.

We now turn to the effect of higher derivative corrections in the EFT. In particular, we would like to determine the value of the parity-odd coefficient $\kappa$ in perturbation theory. We found in Sec.~\ref{ssec_EFTsphere} that $\kappa$ controls the number of vortices (or charges, in the dual picture) in the ground state when the theory is placed on a sphere. As we now show, this phenomenon indeed happens in monopole states in the CFT \eqref{eq_S_anyon}, which will allow us to fix $\kappa$.

To leading order in $1/k$, we can treat $a_\mu$ as a background field. The constraint \eqref{eq_anyonconstraint} implies that the state consists of $N_\psi = kQ$ Dirac fermions on the sphere in a monopole background with flux $Q$. The fermions will populate monopole harmonics with degeneracy%
	\footnote{Landau levels for Dirac fermions have one more available state than regular monopole harmonics \cite{Wu:1976ge}, which have spin $Q/2+p$ and degeneracy $Q+2p +1$, see e.g. \cite{Golkar:2014wwa,Chester:2017vdh}.} 
$Q + 2|p|$, with $p\in \mathbb Z$.  Filling up the levels $p=1,2,\ldots,k$ therefore requires $kQ + k(k+1)$ fermions. The missing fermions (or holes) in the $k$th Landau level are precisely the vortices that were expected in the EFT. Their number fixes the parity-odd coefficient:
\begin{equation}\label{eq_kappaanyons}
\kappa = k(k+1)\, .
\end{equation}
This result can also be found in Ref.~\cite{Golkar:2014wwa}; note that $\kappa$ is even, as expected in CFTs with no gauge invariant fermionic operators. With a microscopic model in hand, we can investigate the internal structure of the EFT vortices (cf.~comments at the end of Sec.~\ref{sec_EFT}): since particles in higher Landau levels have angular momentum $J>Q/2$, we find that the vortices here will carry non-zero internal spin $s=J-Q/2\sim k$.

The large number of vortices $\kappa \gg 1$ implies a rich low-lying spectrum of soft vortex excitations above the ground state. In particular,the ground state will form a vortex lattice as discussed in Sec.~\ref{ssec_Vortex_Lattice}. Eq. \eqref{eq_DeltaQ_vort_largeK} gives the energy of the monopole operator. 
The lightest excitation is instead provided by fluctuations of the vortex coordinates in eq. \eqref{eq_Delta_vortex_lattice_excitations}, with energy scaling as $\Delta-\Delta_{\rm min }(Q)\sim\sqrt{J/kQ}$ for $J\ll k$.

The fermionic nature of the vortices has a striking consequence for the theory at hand. Indeed we find that when $\kappa>Q+1$ they will spill out to the next Landau level which is not captured by the EFT, as discussed in Sec.~\ref{ssec_EFTsphere}. The EFT \eqref{eq_Sfinal} therefore only correctly captures operators of large charge $Q$ satisfying 
\begin{align}
\hbox{EFT regime:} & &
	Q \gtrsim \kappa \simeq k^2\, . & & \phantom{\hbox{EFT regime}}
\end{align}
This window is much smaller than the one in which the superfluid EFT applies in flat space, where the gap of the heavy states scales like $\sqrt{\rho/k}$. It is then natural to wonder if there is an alternative low energy description for $k^2\gtrsim Q\gg k$. To answer this question we will make use of the weak coupling of this CFT.

We work to leading order in $k$ in the following for simplicity. Now from the perspective of the CFT, it is clear what happens when $Q$ is taken below this threshold: the $\kappa\simeq k^2$ missing electrons completely deplete the 
$k$th Landau level, and start depleting the $(k-1)$th Landau level.
 Since the Landau levels have degeneracy $\simeq Q$, the number of filled Landau levels is now no longer $k$, but rather
\begin{equation}
k - \left\lfloor \frac{k^2}{Q}\right\rfloor \, .
\end{equation}
Integrating out the gapped Landau levels now no longer cancels the Chern-Simons term:
\begin{equation}\label{eq_SEFTprime}
S_{\rm EFT'}=
	-\frac1{4\pi}\left\lfloor \frac{k^2}{Q}\right\rfloor \int ada
	+ S_{\rm EFT}\, , 
\end{equation}
where $S_{\rm EFT}$ is still given by \eqref{eq_Sfinal} with coefficients $\alpha,\, \kappa$ unchanged to this level of precision (we are assuming that number of depleted Landau levels is small $\left\lfloor \frac{k^2}{Q}\right\rfloor \ll k$). However, the number of vortices in $S_{\rm EFT}$ is now no longer given by the total number of holes $\kappa\simeq k^2$, but rather by the number of holes in the partially filled Landau level
\begin{equation}\label{eq_Nvortex_expect}
N_{\rm vortices} = k^2 - \left\lfloor \frac{k^2}{Q}\right\rfloor Q\, .
\end{equation}
How is this consistent with Eq.~\eqref{eq_noncons2} for the nonconservation of the higher-form current, which ties the number of vortices to $\kappa$? The resolution is that the Chern-Simons term also breaks the higher-form symmetry, and therefore gives an additional contribution to the divergence of the current
\begin{equation}
\nabla_\mu J^{\mu\nu}
	= -\left\lfloor \frac{k^2}{Q}\right\rfloor j^\nu + \kappa J^\nu_{\rm Euler} - j^\nu_{\rm vortices}
\end{equation}
Integrating the $\nu=0$ component over the spatial sphere gives  \eqref{eq_Nvortex_expect}. We therefore find that the upgraded EFT \eqref{eq_SEFTprime} with Chern-Simons term extends the regime of validity of the EFT to
\begin{align}
\hbox{EFT$'$ regime:} & &
	k^2 \gtrsim Q \gg k\, . & & \phantom{\hbox{EFT regime}}
\end{align}
The physics in this regime is very similar. For instance, the spectrum of phonons is unaffected to leading order by the Chern-Simons term. In flat space, this would give them a gap $m_{\rm phonon}\sim \sqrt{\frac{k}{\rho}}$; however, this is much smaller than $1/R$ and it is hence \emph{invisible} on the sphere. The main difference is the reduced number of vortices in the ground state \eqref{eq_Nvortex_expect}. This affects the spectrum of their excitations as well as the coefficient in front of the fugacity term in the scaling dimension of the monopole operator \eqref{eq_DeltaQ_vort}.

Finally, for $Q/k\lesssim 1$ there is no parametric separation between the gap of the particle-hole excitations and the compactification scale $1/R$, hence the EFT breaks down. As explained around Eq.~\eqref{eq_Deltadouble}, it may still be possible to exploit the weak coupling of the theory to explore this regime. We leave this task for future work.

%======================================================================%
%======================================================================%
%======================================================================%
\subsection{Further applications and prospects}

%======================================================================%
\subsubsection*{Large $N_c$ Chern-Simons matter theories}

Chern-Simons theories with fundamental fermion matter at infinite $N_c$ and finite chemical potential have the thermodynamic properties of a (renormalized) Fermi liquid \cite{Geracie:2015drf,Minwalla:2020ysu}%
	\footnote{In \cite{Berkooz:2006wc,Berkooz:2008gc} it was also suggested that Fermi liquid states in $\mathcal{N}=4$ SYM at infinite $N_c$ might be dual to large charge extremal black holes in $AdS_5$ . 
	}
, even though there are no gauge-invariant fermionic operators for $N_c$ even. However, these are unlikely to be the true ground states at any finite but large $N_c$. Indeed, $1/N_c$ corrections in the presence of a Fermi surface may be more relevant than the terms leading in $N_c$, changing the physics at low energies  (similar effects even spoil large $N_f$ expansions in the presence of a Fermi surface \cite{Lee:2009epi}). Consider for example the gluon propagator, which in axial gauges receives no loop correction at leading order in $1/N_c$ and takes the form $G(p) \sim \frac{1}{p}$ \cite{Giombi:2011kc}. At order $1/N_c$, it receives a self-energy correction from the Fermi surface (Landau damping) 
\begin{equation}
G(p) \sim \frac{1}{p + \Pi(p)}\, , \qquad \hbox{with} \quad
	\Pi(p)\sim \frac{\mu}{N_c} \frac{p_0}{p}\, .
\end{equation}
We expect gauge invariant observables to receive similar corrections, leading to a breakdown of the large $N$ solution at energies $p\lesssim \mu/N_c$. It may be possible to reorganize the perturbative expansion (see for example \cite{Damia:2019bdx}) and establish whether the true ground state is a superfluid or a non-Fermi liquid; we leave this for future work. IR instabilities due to quantum effects may also plague near extremal black holes, and lead to a qualitatively different ground state.

%======================================================================%
\subsubsection*{Non-relativistic CFTs}

It was suggested in a non-relativistic context in Ref.~\cite{Kravec:2018qnu} that superfluid EFTs may describe the large charge sector in scale invariant theories of anyons. We believe the superfluid EFT can indeed be derived in these situations as well following arguments similar to those outlined in this section, see in particuliar Ref.~\cite{Du:2020gqf}. A term similar to $\kappa$ also appears in non-relativistic superfluids, see e.g.~Refs.~\cite{Hoyos:2013eha,Moroz:2015cft}. 

%======================================================================%
%======================================================================%
%======================================================================%
\subsection*{Acknowledgements}

We thank Clay C\'ordova, Angelo Esposito, Ilya Esterlis, Andrey Gromov, Zohar Komargodski, Emil Martinec, M\'ark Mezei, Shiraz Minwalla, Dung X.~Nguyen, Dam T.~Son and Paul Wiegmann for inspiring discussions (alas mostly virtual). We also thank Domenico Orlando for comments on an earlier version of the manuscript. UM is partially supported by NSF grants No. PHY1720480 and PHY201419 and by a Simons Investigator grant from the Simons foundation (PI: Dam Thanh Son). LVD is supported by the Swiss National Science Foundation and the Robert R.~McCormick Postdoctoral Fellowship of the Enrico Fermi Institute. GC is supported by the Simons Foundation (Simons Collaboration on the Non-perturbative Bootstrap) grants 488647 and 397411.

\appendix

%######################################################################%
%======================================================================%
%======================================================================%
%======================================================================%
%######################################################################%
\section{General parity-violating superfluids}\label{app_EFT}

We review in this section the effective field theory for superfluids in 3 spacetime dimensions without conformal symmetry, following Ref.~\cite{Golkar:2014paa}. Similar EFTs have been studied in the non-relativistic context of chiral superfluids, see e.g.~\cite{Hoyos:2013eha,Moroz:2015cft}. As in the main text, we work in the dual formalism where the action is a function of an abelian gauge field $a_\mu$. The stress tensor is then computed, and compared to the one obtained in Ref.~\cite{Jensen:2011xb} from a hydrodynamic approach.

%======================================================================%
%======================================================================%
%======================================================================%
\subsection{General EFT without conformal symmetry}\label{sapp_geneft}

The most general action for parity violating relativistic superfluids in 3 spacetime dimensions was constructed in \cite{Golkar:2014paa}, which we reproduce here
\begin{equation}\label{eq_general_action}
	\begin{split}
		S = ~ &S_\alpha + S_\zeta + S_\kappa + \cdots\\
		= ~ & \sigma \int d^3 x ~ \sqrt{\sigma g} ~ \alpha(|f|)\\
		&+ \int d^3 x ~ \sqrt{\sigma g} ~ \zeta(|f|) \epsilon^{\mu\nu\lambda} u_\mu \d_\nu u_\lambda\\
		&+ \frac{\kappa}{8\pi} \int d^3 x ~ \sqrt{\sigma g} ~ a_\mu \epsilon^{\mu\nu\rho} \epsilon^{\alpha\beta\gamma} u_\alpha \left( \nabla_\nu u_\beta \nabla_\rho u_\gamma + \frac{\sigma}{2} \mathcal{R}_{\nu\rho\beta\gamma} \right) + \cdots\, , 
	\end{split}
\end{equation}
where $\sigma$ is the signature of the metric, and we have defined $f^\mu = \epsilon^{\mu\nu\lambda}\d_\nu a_\lambda$, $|f| = \sqrt{\sigma f_\mu f^\mu}$ and $u_\mu = f_\mu / |f|$ is a unit vector. As in the main text, $\mathcal{R}_{\nu\rho\beta\gamma}$ is the Riemann tensor of the spacetime manifold that the superfluid lives on. Without conformal symmetry $\alpha(|f|)$ and $\zeta(|f|)$ are generic functions of $|f|$; in the conformal case in equation \eqref{eq_aEFT} they were given by
\begin{equation}
	\alpha(|f|) = \alpha \cdot |f|^{3/2}, \qquad \zeta (|f|) = \zeta \cdot |f|,
\end{equation}
where $\alpha$ and $\zeta$ are now constants. The last term in \eqref{eq_general_action} is automatically scale invariant, and has no generalization to the non-conformal case since it is constructed from the identically conserved `Euler' current 
\begin{equation}
	J_E^\mu = \frac{1}{8\pi} \epsilon^{\mu\nu\rho} \epsilon^{\alpha\beta\gamma} u_\alpha \left( \nabla_\nu u_\beta \nabla_\rho u_\gamma + \sigma \frac{1}{2} \mathcal{R}_{\nu\rho\beta\gamma} \right).
\end{equation}
On Euclidean manifolds with a spatial factor of $S^2$, the total charge $\int d^2 x ~ J^0$ measures the winding number of the map $u^\mu:S^2 \rightarrow S^2$, and is hence an integer. Invariance under large gauge transformations imposes $\kappa\in \mathbb Z$ as in the conformal case.

The $\zeta$ term can be eliminated in the conformal case by the field redefinition \eqref{eq_fieldredef}, so it is natural to ask if the same can be done in the general case. The equation of motion from the leading term in the action is
\begin{equation}
	\epsilon^{\mu\nu\lambda} \d_\mu \left( \frac{\alpha'(|f|)f_\nu}{|f|} \right) = 0.
\end{equation}
So a field redefinition of the type
\begin{equation}
	a_\mu \rightarrow a_\mu + \delta a_\mu, \qquad \delta a_\mu = c \frac{\alpha'(|f|) f_\mu}{|f|},
\end{equation}
would leave local gauge invariant observables unchanged on-shell, since
\begin{equation}
	\delta f^\mu = \epsilon^{\mu\nu\lambda} \d_\nu \delta a_\lambda = c \epsilon^{\mu\nu\lambda} \d_\mu \left( \frac{\alpha'(|f|)f_\nu}{|f|} \right) = 0.
\end{equation}
The change in the leading term in the action under this field redefinition is
\begin{equation}
	\delta S_\alpha = - c \int d^3 x ~ \sqrt{-g} (\alpha')^2 \epsilon^{\mu\nu\lambda} u_\mu \d_\nu u_\lambda.
\end{equation}
Hence, if $(\alpha')^2 (|f|) \propto \zeta(|f|)$ as a function, this field redefinition can be used to eliminate the $\zeta$ term altogether by setting
\begin{equation}
	c \equiv \frac{\zeta}{(\alpha')^2}
\end{equation}
In the conformal case $\alpha(|f|) = \alpha |f|^{3/2}$ and $\zeta(|f|) = \zeta |f|$, so this is indeed true and we find $c = 2\alpha/3\zeta$ as in \eqref{eq_fieldredef}. However, $\zeta/(\alpha')^2$ isn't a constant in the general case and the $\zeta$ term cannot be eliminated, even perturbatively.

%======================================================================%
%======================================================================%
%======================================================================%
\subsection{Gravitational stress tensor}\label{sapp_Tmunu}

We compute the gravitational stress tensor for the superfluid EFT in the general, non-conformal case using the action \eqref{eq_general_action}. For simplicity, we restrict ourselves to Euclidean signature. The gravitational variation of the leading order term is straightforward and results in the expression
\begin{equation}
	\begin{split}
		\delta S_\alpha &= \frac{1}{2} \int \sqrt{ g} \left[ \left( \alpha(|f|) - |f| \alpha'(|f|) \right) g^{\mu\nu} + |f| \alpha'(|f|) u^\mu u^\nu \right] \delta g_{\mu\nu}\\
		&= - \frac{1}{2} \int \sqrt{g} ~ T^{\mu\nu} \delta g_{\mu\nu}
	\end{split}
\end{equation}
from which we can read off its contribution to the gravitational stress-tensor to find
\begin{equation}
	T_\alpha^{\mu\nu} = |f| \alpha'(|f|) u^\mu u^\nu + \left( \alpha(|f|) - |f| \alpha'(|f|) \right) g^{\mu\nu}
\end{equation}
In the conformal case $\alpha(|f|) = \alpha |f|^{3/2}$ this expression simplifies to
\begin{equation}\label{eq_st_alpha}
	T_\alpha^{\mu\nu}|_\text{conformal} = \frac{3\alpha}{2} |f|^{3/2} \left( u^\mu u^\nu - \frac{1}{3} g^{\mu\nu} \right)\, . 
\end{equation}
Moving on to the general $\zeta$ term we can use the variational identities
\begin{equation}
	\begin{split}
		\delta |f| &= \frac{1}{2} |f| \left( u^\mu u^\nu - g^{\mu\nu} \right) \delta g_{\mu\nu}\\
		\delta u_\alpha &= u^\mu \left( \delta^\nu_\alpha - u^\nu u_\alpha \right) \delta g_{\mu\nu}
	\end{split}
\end{equation}
to simplify the variation down to
\begin{equation}
	\begin{split}
		\delta S_\zeta = -\frac{1}{2} \int \sqrt{g} &\Bigg[ |f|\zeta'(|f|) g^{\mu\nu} udu + (2\zeta(|f|) - |f|\zeta'(|f|)) u^\mu u^\nu udu\\
		& - 4\zeta(|f|) u^\mu \epsilon^{\nu\alpha\beta} \nabla_\alpha u_\beta + 2\zeta'(|f|) u^\mu \epsilon^{\nu\alpha\beta} u_\alpha \nabla_\beta |f| \Bigg] \delta g_{\mu\nu}
	\end{split}
\end{equation}
where $udu$ is short-hand for $\epsilon^{\mu\nu\lambda}u_\mu\partial_\nu u_\lambda$. The contribution of this term to the stress tensor is then given by
\begin{equation}
	\begin{split}
		T_\zeta^{\mu\nu} &= |f| \zeta'(|f|) g^{\mu\nu} udu + \left[ 2\zeta(|f|)-|f|\zeta'(|f|) \right] u^\mu u^\nu udu\\
		&- 4\zeta(|f|) u^{(\mu}\epsilon^{\nu)\alpha\beta} \nabla_\alpha u_\beta + 2\zeta'(|f|) u^{(\mu} \epsilon^{\nu)\alpha\beta} u_\alpha \nabla_\beta |f| 
	\end{split}
\end{equation}
In the conformal case, this simplifies to
\begin{equation}\label{eq_st_zeta}
	\begin{split}
		T_\zeta^{\mu\nu}|_\text{conformal} &= \zeta |f| g^{\mu\nu} udu + \zeta |f| u^\mu u^\nu udu\\
		&- 4\zeta |f| u^{(\mu}\epsilon^{\nu)\alpha\beta} \nabla_\alpha u_\beta + 2\zeta u^{(\mu} \epsilon^{\nu)\alpha\beta} u_\alpha \nabla_\beta |f|
	\end{split}
\end{equation}
Moving on to the Euler term, using the fact that $\delta u^\alpha = - (u^{\alpha}/2)u^\mu u^\nu \delta g_{\mu\nu}$ and $\delta\mathcal{R}^\beta_{\phantom{\beta}\sigma\nu\rho} = \nabla_\nu \delta \Gamma^\beta_{\rho\sigma} - \nabla_\rho \delta \Gamma^\beta_{\nu\sigma}$ we arrive at the following expression for the gravitational variation of the action
\begin{equation}
	\begin{split}
		\frac{8\pi}{\kappa}\delta S_\kappa &= \frac{1}{2} \int \sqrt{g} a_\mu \epsilon^{\mu\nu\rho} \epsilon_{\alpha\beta\gamma} u^\alpha \left( \nabla_\nu u^\beta \nabla_\rho u^\gamma + \frac{1}{2} \mathcal{R}_{\nu\rho}^{\phantom{\nu\rho}\beta\gamma} \right) g^{\alpha\beta} \delta g_{\alpha\beta}\\
		&- \frac{1}{2} \int \sqrt{g} a_\mu \epsilon^{\mu\nu\rho} \epsilon_{\alpha\beta\gamma} u^\alpha \left( 3 \nabla_\nu u^\beta \nabla_\rho u^\gamma + \frac{1}{2} \mathcal{R}_{\nu\rho}^{\phantom{\nu\rho}\beta\gamma} \right) u^\sigma u^\lambda \delta g_{\sigma\lambda}\\
		&- \frac{1}{2} \int \sqrt{g} a_\mu \epsilon^{\mu\nu\rho} \epsilon_{\alpha\beta\gamma} u^\alpha \mathcal{R}_{\nu\rho}^{\phantom{\nu\rho}\beta\lambda} g^{\gamma\sigma} \delta g_{\sigma\lambda}\\
		&+ 2\int \sqrt{g} a_\mu \epsilon^{\mu\nu\rho} \epsilon_{\alpha\beta\gamma} u^\alpha u^\sigma \nabla_\rho u^\gamma \delta \Gamma^\beta_{\nu\sigma}\\
		&+ \int \sqrt{g} |f| \epsilon^{\mu\nu\rho} \epsilon_{\alpha\beta\gamma} u^\alpha g^{\gamma\sigma} u^\nu \delta \Gamma^\beta_{\nu\sigma} \\
		&+ \int \sqrt{g} a_\mu \epsilon^{\mu\nu\rho} \epsilon_{\alpha\beta\gamma} g^{\gamma\sigma} \nabla_\rho u^\alpha \delta \Gamma^\beta_{\nu\sigma}
	\end{split}
\end{equation}
Next, we can simplify the last three lines further using the variational identity
\begin{equation}
	\delta \Gamma^\beta_{\nu\sigma} = -\frac{1}{2} g^{\beta\lambda} \left( \nabla_\nu \delta g_{\beta\sigma} + \nabla_\sigma \delta g_{\beta\nu} - \nabla_\beta \delta g_{\nu\sigma} \right)
\end{equation}
Simplifying further, we find that the variation of the action organizes itself into a sum of two terms
\begin{equation}
	\delta S_\kappa = \delta S_{\kappa,f} + \delta S_{\kappa,a}
\end{equation}
where $\delta S_f$ contains terms depending directly on $f^\mu$ that are local in the dual scalar language, while $\delta S_a$ contains terms that explicitly depend on the gauge field and lead to non-local, gauge dependent contributions to the stress tensor. We find, for the explicitly gauge invariant part
\begin{equation}\label{eq_euler_gvar_f}
	\begin{split}
		\delta S_{\kappa,f} = \frac{\kappa}{8\pi} \int \sqrt{g} &\Bigg( |f| u^\mu \epsilon^{\nu\alpha\beta} u_\alpha (u\cdot \nabla) u_\beta - u^\mu \epsilon^{\nu\alpha\beta} u_\alpha \nabla_\beta |f|\\
		&+ |f| u_\mu \epsilon^{\nu\alpha\beta} \nabla_\alpha u_\beta - |f| u_\alpha \epsilon^{\alpha\beta\mu} \nabla_\beta u^\nu \Bigg) \delta g_{\mu\nu}
	\end{split}
\end{equation}
The gauge dependent part, on the other hand, is given by the following sum
\begin{equation}
	\begin{split}
		\frac{8\pi}{\kappa}\delta S_{\kappa,a} &= \frac{1}{2} \int \sqrt{g} ~ a_\mu \epsilon^{\mu\nu\rho}\epsilon^{\alpha\beta\gamma} u_\alpha\nabla_\nu u_\beta\nabla_\rho u_\gamma \left( g^{\sigma\lambda} - 3u^\sigma u^\lambda \right) \delta g_{\sigma\lambda}\\
		&+ \frac{1}{2} \int \sqrt{g} a_\mu \epsilon^{\mu\nu\rho}\epsilon^{\alpha\beta\gamma} u_\alpha \mathcal{R}_{\nu\rho\beta\gamma} \left( g^{\sigma\lambda} - u^\sigma u^\lambda \right) \delta g_{\sigma\lambda}\\
		&- \frac{1}{2} \int \sqrt{g} a_\mu \epsilon^{\mu\nu\rho}\epsilon^{\alpha\beta\sigma} u_\alpha \mathcal{R}_{\nu\rho\beta}^{\phantom{\nu\rho\beta}\lambda} \delta g_{\sigma\lambda}\\
		&- \int \sqrt{g} \epsilon^{\mu\nu\rho} \epsilon^{\alpha\beta\gamma} a_\mu \nabla_\nu [ (P_\parallel)^\sigma_\gamma \nabla_\rho u_\alpha ] \delta g_{\beta\sigma}\\
		&+ \int \sqrt{g} \epsilon^{\mu\nu\rho} \epsilon^{\alpha\beta\gamma} a_\mu \nabla_\beta [ (P_\parallel)^\sigma_\gamma \nabla_\rho u_\alpha ] \delta g_{\nu\sigma}\\
		&+ \int \sqrt{g} \epsilon^{\mu\nu\rho} \epsilon^{\alpha\beta\gamma} a_\mu \nabla_\sigma  [ (P_\perp)^\sigma_\gamma \nabla_\rho u_\alpha ] \delta g_{\beta\nu}\\
		&+ \int \sqrt{g} \epsilon^{\mu\nu\rho} \epsilon^{\alpha\beta\gamma} \nabla_\sigma a_\mu (P_\perp)^\sigma_\gamma \nabla_\rho u_\alpha \delta g_{\beta\nu}\\
		&+ \int \sqrt{g} \epsilon^{\mu\nu\rho} \epsilon^{\alpha\beta\gamma} \nabla_\beta a_\mu (P_\parallel)^\sigma_\gamma \nabla_\rho u_\alpha \delta g_{\sigma\nu}
	\end{split}
\end{equation}
where $P_\parallel^{\mu\nu} = u^\mu u^\nu$ and $P_\perp^{\mu\nu} = g^{\mu\nu} - u^\mu u^\nu$. To simplify these further, we use the following strategy: each factor with a lower $\alpha, \beta$ or $\gamma$ index is multiplied by a Kronecker delta expanded in projectors parallel and transverse to $u$. Since the product of these factors is multiplied by $\epsilon^{\alpha\beta\gamma}$, the only terms that survive upon expanding are those that have exactly one factor of $P_\parallel$ and two factors of $P_\perp$, by virtue of the properties of scalar triple products. We also use the fact that $(P_\parallel)^\lambda_\alpha \nabla_\delta u_\lambda = 0$. This trick can be used, for instance, to show that the last two lines in the expression above cancel with each other. Finally, we arrive at the following expression
\begin{equation}\label{eq_nongauge}
	\begin{split}
		\frac{8\pi}{\kappa}\delta S_{\kappa,a} &= \frac{1}{2} \int \sqrt{g} a_\mu \epsilon^{\mu\nu\rho}\epsilon^{\alpha\beta\gamma} u_\alpha \mathcal{R}_{\nu\rho\beta\gamma} \left( g^{\sigma\lambda} - u^\sigma u^\lambda \right) \delta g_{\sigma\lambda}\\
		&- \frac{1}{2} \int \sqrt{g} a_\mu \epsilon^{\mu\nu\rho}\epsilon^{\alpha\beta\sigma} u_\alpha \mathcal{R}_{\nu\rho\beta\delta} \left( g^{\delta\lambda} - u^\delta u^\lambda \right) \delta g_{\sigma\lambda}\\
		&+ \frac{1}{2} \int \sqrt{g}  a_\mu \epsilon^{\mu\nu\rho}\epsilon^{\alpha\beta\gamma} u_\alpha\nabla_\nu u_\beta\nabla_\rho u_\gamma \left( g^{\sigma\lambda} - 3u^\sigma u^\lambda \right) \delta g_{\sigma\lambda}\\
		&+ \int \sqrt{g} a_\mu \epsilon^{\mu\nu\rho} \epsilon^{\alpha\beta\gamma} u_\alpha \nabla_\beta u^\sigma \nabla_\rho u_\sigma \delta g_{\gamma\nu}\\
		&- \int \sqrt{g} a_\mu \epsilon^{\mu\nu\rho} \epsilon^{\alpha\beta\gamma} \nabla_\rho u_\alpha \left[ \delta^\lambda_\beta \nabla_\nu (P_\parallel)^\sigma_\gamma + \delta^\sigma_\beta \delta^\lambda_\nu \nabla_\delta (P_\parallel)^\delta_\gamma - \delta_\nu^\lambda \nabla_\beta (P_\parallel)^\sigma_\gamma \right] \delta g_{\sigma\lambda}
	\end{split}
\end{equation}
The terms in the integrand that multiply $\delta g$ in the first two lines are linear in Goldstone fluctuations while those in the last three lines are at least quadratic. Only terms linear in the Goldstone are needed to study the transport properties of the superfluid at leading order in small frequencies and momenta in Secs.~\ref{ssec_OPEphonon} and \ref{sapp_T_lin}; we will therefore drop the last three lines. The first two lines give contributions to transport on curved manifolds, such as the sphere at small momenta $k\sim 1/R$, and consequently will affect the OPE coefficients in Sec.~\ref{ssec_OPEphonon} at small spin $J\sim 1$. However, phonons with wavelengths of the order of the sphere radius will be sensitive to the $|\kappa|$ vortices distributed on the sphere -- correlation functions with $k\sim 1/R$ therefore require a careful treatment of the vortex-superfluid system \eqref{eq_Stot}. We expect that such a treatment will also resolve the apparent gauge non-invariance of the stress tensor on the sphere. In the remainder of this section, we focus on the leading stress tensor on the plane, where all non-gauge invariant terms \eqref{eq_nongauge} can be ignored.

From \eqref{eq_euler_gvar_f} we find the gauge invariant part of the stress tensor
\begin{equation}
	\begin{split}
		T_{\kappa,f}^{\mu\nu} = \frac{\kappa}{4\pi} &\Bigg[ - |f| u^{\langle\mu} \epsilon^{\nu\rangle\alpha\beta} u_\alpha (u\cdot\nabla) u_\beta + u^{\langle\mu}\epsilon^{\nu\rangle\alpha\beta} u_\alpha \nabla_\beta |f|\\
		&- |f| u^{\langle\mu} \epsilon^{\nu\rangle\alpha\beta} \nabla_\alpha u_\beta + |f| u_\alpha \epsilon^{\alpha\beta \langle\mu} \nabla_\beta u^{\nu\rangle} \Bigg]
	\end{split}
\end{equation}
where the angular brackets stand for symmetrization and trace subtraction. Putting this together with \eqref{eq_st_alpha} and \eqref{eq_st_zeta}, the gauge invariant part of the stress tensor is given by
\begin{equation}
	\begin{split}
		T^{\mu\nu}_f &= \frac{3\alpha}{2} |f|^{3/2} u^{\langle\mu} u^{\nu\rangle}\\
		&+ \zeta \left( t_0^{\mu\nu} - 4 t_3^{\mu\nu} + 2 t_4^{\mu\nu} \right)\\
		&\frac{\kappa}{4\pi} \left( - t_1^{\mu\nu} + t_4^{\mu\nu} - t_3^{\mu\nu} + t_5^{\mu\nu} \right)
	\end{split}
\end{equation}
in a basis of gauge invariant symmetric, traceless tensors constructed from a single factor of $f$, one derivative and $u_\mu$'s
\begin{equation}
	\begin{split}
		t_0^{\mu\nu} &= |f| u^{\langle\mu} u^{\nu\rangle} \epsilon^{\alpha\beta\gamma}u_\alpha\nabla_\beta u_\gamma\\
		t_1^{\mu\nu} &= |f| u^{\langle\mu} \epsilon^{\nu\rangle\alpha\beta} u_\alpha (u\cdot\nabla) u_\beta\\
		t_2^{\mu\nu} &= |f| u_\alpha \epsilon^{\alpha\beta\langle\mu}\nabla^{\nu\rangle} u_\beta\\
		t_3^{\mu\nu} &= |f| u^{\langle\mu} \epsilon^{\nu\rangle\alpha\beta} \nabla_\alpha u_\beta\\
		t_4^{\mu\nu} &= u^{\langle\mu}\epsilon^{\nu\rangle\alpha\beta} u_\alpha \nabla_\beta |f|\\
		t_5^{\mu\nu} &= |f| u_\alpha \epsilon^{\alpha\beta \langle\mu} \nabla_\beta u^{\nu\rangle}
	\end{split}
\end{equation}
These tensors are not linearly independent and we find the following relations between them
\begin{equation}
	t_3^{\mu\nu} = t_0^{\mu\nu} + t_1^{\mu\nu}, \qquad t_2^{\mu\nu} = t_0^{\mu\nu} + t_1^{\mu\nu} + t_5^{\mu\nu}
\end{equation}
so that the gauge invariant part of the stress tensor becomes
\begin{equation}
	\begin{split}
		T^{\mu\nu}_f &= \frac{3\alpha}{2} |f|^{3/2} u^{\langle\mu} u^{\nu\rangle}\\
		&+ \zeta \left( -3 t_0^{\mu\nu} - 4 t_1^{\mu\nu} + 2 t_4^{\mu\nu} \right)\\
		&\frac{\kappa}{4\pi} \left( - t_0^{\mu\nu} - 2 t_1^{\mu\nu} + t_4^{\mu\nu} + t_5^{\mu\nu} \right)
	\end{split}
\end{equation}
Finally, to leading order in the derivative expansion in the EFT these terms can be simplified further on-shell using the leading equations of motion, whose projections parallel and perpendicular to $u^\mu$ are given, respectively, by
\begin{equation}\label{eq_app_eom}
	\begin{split}	
		udu &= 0\\
		|f| (u\cdot\nabla) u_\alpha &= \frac{1}{2} (P_\perp)_\alpha^\beta \nabla_\beta |f|
	\end{split}
\end{equation}
These result in the on-shell relations
\begin{equation}
	t_0^{\mu\nu} = 0, \qquad t_1^{\mu\nu} = \frac{1}{2}t_4^{\mu\nu}
\end{equation}
We then find that the $\zeta$ term vanishes and the $\kappa$ term simplifies to give
\begin{equation}
	T^{\mu\nu}_f = \frac{3\alpha}{2} |f|^{3/2} u^{\langle\mu} u^{\nu\rangle} + \frac{\kappa}{4\pi} |f| u_\alpha \epsilon^{\alpha\beta\langle\mu} \nabla_\beta u^{\nu\rangle}\,.
\end{equation}

%======================================================================%
%======================================================================%
%======================================================================%
\subsection{Comparison with parity-violating hydrodynamics}

Ref.~\cite{Jensen:2011xb} obtained the most general constitutive relations for a $U(1)$ current and stress tensor of a parity-violating fluid in $2+1$ dimensions, up to first order in gradients. These are local expressions for the currents, in a derivative expansion, in terms of the fluid degrees of freedom: fluctuations in temperature $T$, chemical potential $\mu$ and a velocity vector $u^\mu$ satisfying $u^2 = -1$. These expressions should apply to our superfluid as a special case, with the following restrictions: $T$ is not a degree of freedom in a zero-temperature QFT, dissipative terms such as the bulk and shear viscosities are set to zero, and finally in the superfluid the vorticity vanishes $\epsilon^{\mu\nu\lambda} u_\mu \d_\nu u_\lambda=0$ (we ignore vortices in this section). The constitutive relations of Ref.~\cite{Jensen:2011xb} in Landau frame are then given by
\begin{subequations}
\begin{align}
T_{\mu\nu}
	&= (\epsilon+P) u_\mu u_\nu + P g_{\mu\nu} - \tilde \eta \tilde \sigma_{\mu\nu} + \cdots \, , \\
j^\mu
	&= \rho u^\mu - \tilde \sigma \epsilon^{\mu\nu\lambda}u_\nu \d_\lambda \mu + \cdots\, ,
\end{align}
\end{subequations}
with 
\begin{equation}
\tilde \sigma^{\mu\nu}
	= u^\mu \epsilon^{\nu\alpha\beta} u_\alpha (u\cdot \nabla) u_\beta + u_\alpha \epsilon^{\alpha \beta\mu}\nabla_{\beta}u^\nu + (\mu \leftrightarrow \nu)\, .
\end{equation}
We used $\epsilon^{\mu\nu\lambda} u_\mu \d_\nu u_\lambda=0$ to simplify terms. The theory contains two parity-odd 1-derivative coefficients, the Hall conductivity $\tilde \sigma$ and Hall viscosity $\tilde \eta$. In studying the EFT \eqref{eq_EFT_intro}, we defined the unit vector to point in the direction of the current, so that $j^\mu \equiv \rho u^\mu$. From the perspective of hydrodynamic constitutive relations, this amounts to working in Eckart frame, which can be reached with the redefinition $u^\mu \to u^\mu + \delta u^\mu$ with
\begin{equation}
\delta u^\mu
	= \frac{\tilde\sigma }{\rho} \epsilon^{\mu\nu\lambda}u_\nu \d_\lambda \mu\, .
\end{equation}
After using the thermodynamic identities $\epsilon + P = \mu n$, $d \epsilon = \mu dn$ and the continuity relation to leading order $\d_\mu T^{\mu\nu}$, the 1-derivative stress tensor in Eckart frame can be expressed
\begin{equation}
T^{\mu\nu}
	= (\epsilon + P)\varepsilon u^\mu u^\nu + P g_{\mu\nu} +  2\tilde\eta  u_\alpha \epsilon^{\alpha\beta (\mu}\nabla_\beta u^{\nu)}  + \left(\tilde \eta - \mu^2 \tilde \sigma\right) u^{(\mu} \epsilon^{\nu)\alpha\beta} u_\alpha (u\cdot \nabla) u_\beta  \, ,
\end{equation}
where we denoted symmetrization by $A^{(\mu\nu)}\equiv \frac{1}{2}(A^{\mu\nu} + A^{\nu\mu})$. This agrees with the stress tensor obtained from the EFT \eqref{eq_T}, with
\begin{equation}
\tilde \eta = \frac{\kappa n}{4}\, , \qquad
\tilde \sigma = \frac{\kappa n}{4\mu^2}\, .
\end{equation}
%
%======================================================================%
%======================================================================%
%======================================================================%
\subsection{Linearized stress tensor}\label{sapp_T_lin}

We obtain in this section the matrix elements of the stress tensor in single phonon states, which lead to predictions for OPE coefficients in Sec.~\ref{ssec_OPEphonon}. Linearizing the stress tensor \eqref{eq_T}, after using the field redefinition \eqref{eq_a_redef_pert}, gives
\begin{subequations}\label{eq_T_expand}
\begin{align}
T^{00}	\label{seq_T_exp}
	&= \alpha (2\pi \rho)^{3/2} + \frac{3}{2} \alpha \sqrt{2\pi \rho} \, b\ + \cdots\, , \\
T^{0i}
	&= \frac{3}{2} \alpha \sqrt{2\pi \rho}(-2\pi j^i) + \cdots\, , \\
T^{ij}
	&= \frac{1}{2}\delta^{ij} \alpha (2\pi \rho)^{3/2} + \frac{3}{4}\delta^{ij} \alpha \sqrt{2\pi \rho} \,b + \frac{\kappa}{8\pi} \left(\nabla^i e^j + \nabla^j e^i\right) + \cdots\,,
\end{align}
\end{subequations}
with $j^i$ given by \eqref{eq_j_expand}.
Using the mode expansion \eqref{eq_mode} and \eqref{eq_eb_to_pic} then leads to the following matrix elements
\begin{equation}
	\begin{split}
		\langle Q | T_{\mu\nu} | Q,Jm \rangle_{\rm even} = ~ &i \left(\frac{\rho^3}{8\chi_0}\right)^{1/4} \sqrt{\Omega_J}\left( 3\delta_\mu^t \delta_\nu^t + g_{\mu\nu} \right) {e^{-i\Omega_J t}} Y_{Jm}(\hat{n})\\
		&+ \left(\frac{\rho^3}{8\chi_0}\right)^{1/4} \frac1{\sqrt{\Omega_J}} \left( \delta_\mu^i \delta_\nu^t + \delta_\nu^i \delta_\mu^t \right) e^{-i\Omega_J t} \partial_i Y_{Jm}(\hat{n})
	\end{split}
\end{equation}
for the parity-even part, and 
\begin{equation}
	\begin{split}
		\langle Q | T^{\mu\nu} | Q,Jm \rangle_{\rm odd} &= i\frac{\kappa}{4} \left(\frac{\rho\chi_0}{2}\right)^{1/4}  \sqrt{\Omega_J} \left( \delta^\mu_t \delta^\nu_i + \delta^\nu_t \delta^\mu_i \right)  e^{-i\Omega_J t} \epsilon^{ij} \partial_jY_{Jm}(\hat{n}) \\
		&- \frac{\kappa}{8} \left(\frac{\rho\chi_0}{2}\right)^{1/4}  \frac1{\sqrt{\Omega_J}} \left( g^{\mu i} \delta^\nu_j + g^{\nu i} \delta^\mu_j \right) \epsilon^{jk} \nabla_i \partial_k e^{-i\Omega_J t}Y_{Jm} (\hat{n})
	\end{split}
\end{equation}
for the parity-odd part of the stress tensor.

\bibliographystyle{ourbst}
\bibliography{parity_Q}{}

\providecommand{\href}[2]{#2}\begingroup\raggedright\begin{thebibliography}{10}

\bibitem{Alday:2007mf}
L.~F. Alday and J.~M. Maldacena, {{Comments on operators with large spin}},
  \href{http://dx.doi.org/10.1088/1126-6708/2007/11/019}{JHEP {\bf 11}, 019,
  2007}, [\href{http://arxiv.org/abs/arXiv:0708.0672}{{arXiv:0708.0672
  [hep-th]}}].

\bibitem{Fitzpatrick:2012yx}
A.~L. Fitzpatrick, J.~Kaplan, D.~Poland and D.~Simmons-Duffin, {{The Analytic
  Bootstrap and AdS Superhorizon Locality}},
  \href{http://dx.doi.org/10.1007/JHEP12(2013)004}{JHEP {\bf 12}, 004, 2013},
  [\href{http://arxiv.org/abs/arXiv:1212.3616}{{arXiv:1212.3616 [hep-th]}}].

\bibitem{Komargodski:2012ek}
Z.~Komargodski and A.~Zhiboedov, {{Convexity and Liberation at Large Spin}},
  \href{http://dx.doi.org/10.1007/JHEP11(2013)140}{JHEP {\bf 11}, 140, 2013},
  [\href{http://arxiv.org/abs/arXiv:1212.4103}{{arXiv:1212.4103 [hep-th]}}].

\bibitem{Caron-Huot:2017vep}
S.~Caron-Huot, {{Analyticity in Spin in Conformal Theories}},
  \href{http://dx.doi.org/10.1007/JHEP09(2017)078}{JHEP {\bf 09}, 078, 2017},
  [\href{http://arxiv.org/abs/arXiv:1703.00278}{{arXiv:1703.00278 [hep-th]}}].

\bibitem{Hellerman:2015nra}
S.~Hellerman, D.~Orlando, S.~Reffert and M.~Watanabe, {{On the CFT Operator
  Spectrum at Large Global Charge}},
  \href{http://dx.doi.org/10.1007/JHEP12(2015)071}{JHEP {\bf 12}, 071, 2015},
  [\href{http://arxiv.org/abs/arXiv:1505.01537}{{arXiv:1505.01537 [hep-th]}}].

\bibitem{Monin:2016jmo}
A.~Monin, D.~Pirtskhalava, R.~Rattazzi and F.~K. Seibold, {{Semiclassics,
  Goldstone Bosons and CFT data}},
  \href{http://dx.doi.org/10.1007/JHEP06(2017)011}{JHEP {\bf 06}, 011, 2017},
  [\href{http://arxiv.org/abs/arXiv:1611.02912}{{arXiv:1611.02912 [hep-th]}}].

\bibitem{Jafferis:2017zna}
D.~Jafferis, B.~Mukhametzhanov and A.~Zhiboedov, {{Conformal Bootstrap At Large
  Charge}}, \href{http://dx.doi.org/10.1007/JHEP05(2018)043}{JHEP {\bf 05},
  043, 2018}, [\href{http://arxiv.org/abs/arXiv:1710.11161}{{arXiv:1710.11161
  [hep-th]}}].

\bibitem{Hellerman:2017sur}
S.~Hellerman and S.~Maeda, {{On the Large $R$-charge Expansion in ${\mathcal N}
  = 2$ Superconformal Field Theories}},
  \href{http://dx.doi.org/10.1007/JHEP12(2017)135}{JHEP {\bf 12}, 135, 2017},
  [\href{http://arxiv.org/abs/arXiv:1710.07336}{{arXiv:1710.07336 [hep-th]}}].

\bibitem{Lashkari:2016vgj}
N.~Lashkari, A.~Dymarsky and H.~Liu, {{Eigenstate Thermalization Hypothesis in
  Conformal Field Theory}},
  \href{http://dx.doi.org/10.1088/1742-5468/aab020}{J. Stat. Mech. {\bf 1803},
  033101, 2018},
  [\href{http://arxiv.org/abs/arXiv:1610.00302}{{arXiv:1610.00302 [hep-th]}}].

\bibitem{Alday:2019qrf}
L.~F. Alday and E.~Perlmutter, {{Growing Extra Dimensions in AdS/CFT}},
  \href{http://dx.doi.org/10.1007/JHEP08(2019)084}{JHEP {\bf 08}, 084, 2019},
  [\href{http://arxiv.org/abs/arXiv:1906.01477}{{arXiv:1906.01477 [hep-th]}}].

\bibitem{Delacretaz:2020nit}
L.~V. Delacretaz, {{Heavy Operators and Hydrodynamic Tails}},
  \href{http://dx.doi.org/10.21468/SciPostPhys.9.3.034}{SciPost Phys. {\bf 9},
  034, 2020}, [\href{http://arxiv.org/abs/arXiv:2006.01139}{{arXiv:2006.01139
  [hep-th]}}].

\bibitem{Belin:2020hea}
A.~Belin and J.~de~Boer, {{Random Statistics of OPE Coefficients and Euclidean
  Wormholes}},  2020,
  [\href{http://arxiv.org/abs/arXiv:2006.05499}{{arXiv:2006.05499 [hep-th]}}].

\bibitem{Golkar:2014paa}
S.~Golkar, M.~M. Roberts and D.~T. Son, {{The Euler current and relativistic
  parity odd transport}}, \href{http://dx.doi.org/10.1007/JHEP04(2015)110}{JHEP
  {\bf 04}, 110, 2015},
  [\href{http://arxiv.org/abs/arXiv:1407.7540}{{arXiv:1407.7540 [hep-th]}}].

\bibitem{PhysRevLett.60.2677}
R.~B. Laughlin, {Superconducting ground state of noninteracting particles
  obeying fractional statistics},
  \href{http://dx.doi.org/10.1103/PhysRevLett.60.2677}{Phys. Rev. Lett. {\bf
  60}, 2677--2680, 1988}.

\bibitem{Chen:1989xs}
Y.-H. Chen, F.~Wilczek, E.~Witten and B.~I. Halperin, {{On Anyon
  Superconductivity}}, \href{http://dx.doi.org/10.1142/S0217979289000725}{Int.
  J. Mod. Phys. {\bf B3}, 1001, 1989}.

\bibitem{Cuomo:2017vzg}
G.~Cuomo, A.~de~la Fuente, A.~Monin, D.~Pirtskhalava and R.~Rattazzi,
  {{Rotating superfluids and spinning charged operators in conformal field
  theory}}, \href{http://dx.doi.org/10.1103/PhysRevD.97.045012}{Phys. Rev. D
  {\bf 97}, 045012, 2018},
  [\href{http://arxiv.org/abs/arXiv:1711.02108}{{arXiv:1711.02108 [hep-th]}}].

\bibitem{Goon:2014ika}
G.~Goon, A.~Joyce and M.~Trodden, {{Spontaneously Broken Gauge Theories and the
  Coset Construction}},
  \href{http://dx.doi.org/10.1103/PhysRevD.90.025022}{Phys. Rev. D {\bf 90},
  025022, 2014}, [\href{http://arxiv.org/abs/arXiv:1405.5532}{{arXiv:1405.5532
  [hep-th]}}].

\bibitem{Goon:2012dy}
G.~Goon, K.~Hinterbichler, A.~Joyce and M.~Trodden, {{Galileons as Wess-Zumino
  Terms}}, \href{http://dx.doi.org/10.1007/JHEP06(2012)004}{JHEP {\bf 06}, 004,
  2012}, [\href{http://arxiv.org/abs/arXiv:1203.3191}{{arXiv:1203.3191
  [hep-th]}}].

\bibitem{Delacretaz:2014jka}
L.~V. Delacr\'etaz, A.~Nicolis, R.~Penco and R.~A. Rosen, {{Wess-Zumino Terms
  for Relativistic Fluids, Superfluids, Solids, and Supersolids}},
  \href{http://dx.doi.org/10.1103/PhysRevLett.114.091601}{Phys. Rev. Lett. {\bf
  114}, 091601, 2015},
  [\href{http://arxiv.org/abs/arXiv:1403.6509}{{arXiv:1403.6509 [hep-th]}}].

\bibitem{Jensen:2011xb}
K.~Jensen, M.~Kaminski, P.~Kovtun, R.~Meyer, A.~Ritz and A.~Yarom,
  {{Parity-Violating Hydrodynamics in 2+1 Dimensions}},
  \href{http://dx.doi.org/10.1007/JHEP05(2012)102}{JHEP {\bf 05}, 102, 2012},
  [\href{http://arxiv.org/abs/arXiv:1112.4498}{{arXiv:1112.4498 [hep-th]}}].

\bibitem{PhysRevB.86.245309}
B.~Bradlyn, M.~Goldstein and N.~Read, {{Kubo formulas for viscosity: Hall
  viscosity, Ward identities, and the relation with conductivity}},
  \href{http://dx.doi.org/10.1103/PhysRevB.86.245309}{Phys. Rev. B {\bf 86},
  245309, 2012}, [\href{http://arxiv.org/abs/arXiv:1207.7021}{{arXiv:1207.7021
  [cond-mat.stat-mech]}}].

\bibitem{Read:1999fn}
N.~Read and D.~Green, {{Paired states of fermions in two-dimensions with
  breaking of parity and time reversal symmetries, and the fractional quantum
  Hall effect}}, \href{http://dx.doi.org/10.1103/PhysRevB.61.10267}{Phys. Rev.
  B {\bf 61}, 10267, 2000},
  [\href{http://arxiv.org/abs/arXiv:cond-mat/9906453}{{arXiv:cond-mat/9906453}}].

\bibitem{Moroz:2015cft}
S.~Moroz, C.~Hoyos and L.~Radzihovsky, {{Chiral $p\pm ip$ superfluid on a
  sphere}}, \href{http://dx.doi.org/10.1103/PhysRevB.93.024521}{Phys. Rev. {\bf
  B93}, 024521, 2016},
  [\href{http://arxiv.org/abs/arXiv:1511.03502}{{arXiv:1511.03502
  [cond-mat.quant-gas]}}].

\bibitem{Gromov:2014gta}
A.~Gromov and A.~G. Abanov, {{Density-curvature response and gravitational
  anomaly}}, \href{http://dx.doi.org/10.1103/PhysRevLett.113.266802}{Phys. Rev.
  Lett. {\bf 113}, 266802, 2014},
  [\href{http://arxiv.org/abs/arXiv:1403.5809}{{arXiv:1403.5809
  [cond-mat.str-el]}}].

\bibitem{CAN2015752}
T.~Can, M.~Laskin and P.~B. Wiegmann, {Geometry of quantum hall states:
  Gravitational anomaly and transport coefficients},
  \href{http://dx.doi.org/https://doi.org/10.1016/j.aop.2015.02.013}{Annals of
  Physics {\bf 362}, 752--794, 2015},
  [\href{http://arxiv.org/abs/arXiv:1411.3105}{{arXiv:1411.3105
  [cond-mat.str-el]}}].

\bibitem{PhysRevLett.69.953}
X.~G. Wen and A.~Zee, {Shift and spin vector: New topological quantum numbers
  for the hall fluids},
  \href{http://dx.doi.org/10.1103/PhysRevLett.69.953}{Phys. Rev. Lett. {\bf
  69}, 953--956, 1992}.

\bibitem{PhysRevB.84.085316}
N.~Read and E.~H. Rezayi, {{Hall viscosity, orbital spin, and geometry: paired
  superfluids and quantum Hall systems}},
  \href{http://dx.doi.org/10.1103/PhysRevB.84.085316}{Phys. Rev. B {\bf 84},
  085316, 2011}, [\href{http://arxiv.org/abs/arXiv:1008.0210}{{arXiv:1008.0210
  [cond-mat.mes-hall]}}].

\bibitem{Golkar:2014wwa}
S.~Golkar, M.~M. Roberts and D.~T. Son, {{Effective Field Theory of
  Relativistic Quantum Hall Systems}},
  \href{http://dx.doi.org/10.1007/JHEP12(2014)138}{JHEP {\bf 12}, 138, 2014},
  [\href{http://arxiv.org/abs/arXiv:1403.4279}{{arXiv:1403.4279
  [cond-mat.mes-hall]}}].

\bibitem{Horn:2015zna}
B.~Horn, A.~Nicolis and R.~Penco, {{Effective string theory for vortex lines in
  fluids and superfluids}},
  \href{http://dx.doi.org/10.1007/JHEP10(2015)153}{JHEP {\bf 10}, 153, 2015},
  [\href{http://arxiv.org/abs/arXiv:1507.05635}{{arXiv:1507.05635 [hep-th]}}].

\bibitem{Wu:1976ge}
T.~T. Wu and C.~N. Yang, {{Dirac Monopole Without Strings: Monopole
  Harmonics}}, \href{http://dx.doi.org/10.1016/0550-3213(76)90143-7}{Nucl.
  Phys. B {\bf 107}, 365, 1976}.

\bibitem{PhysRevD.41.661}
G.~V. Dunne, R.~Jackiw and C.~A. Trugenberger, {"topological" (chern-simons)
  quantum mechanics}, \href{http://dx.doi.org/10.1103/PhysRevD.41.661}{Phys.
  Rev. D {\bf 41}, 661--666, 1990}.

\bibitem{DUNNE1993114}
G.~Dunne and R.~Jackiw, {“peierls substitution” and chern-simons quantum
  mechanics},
  \href{http://dx.doi.org/https://doi.org/10.1016/0920-5632(93)90376-H}{Nuclear
  Physics B - Proceedings Supplements {\bf 33}, 114 -- 118, 1993}.

\bibitem{Cuomo:2019ejv}
G.~Cuomo, {{Superfluids, vortices and spinning charged operators in 4d CFT}},
  \href{http://dx.doi.org/10.1007/JHEP02(2020)119}{JHEP {\bf 02}, 119, 2020},
  [\href{http://arxiv.org/abs/arXiv:1906.07283}{{arXiv:1906.07283 [hep-th]}}].

\bibitem{Banerjee:2017fcx}
D.~Banerjee, S.~Chandrasekharan and D.~Orlando, {{Conformal dimensions via
  large charge expansion}},
  \href{http://dx.doi.org/10.1103/PhysRevLett.120.061603}{Phys. Rev. Lett. {\bf
  120}, 061603, 2018},
  [\href{http://arxiv.org/abs/arXiv:1707.00711}{{arXiv:1707.00711 [hep-lat]}}].

\bibitem{Iliesiu:2018fao}
L.~Iliesiu, M.~Kolo\u{g}lu, R.~Mahajan, E.~Perlmutter and D.~Simmons-Duffin,
  {{The Conformal Bootstrap at Finite Temperature}},
  \href{http://dx.doi.org/10.1007/JHEP10(2018)070}{JHEP {\bf 10}, 070, 2018},
  [\href{http://arxiv.org/abs/arXiv:1802.10266}{{arXiv:1802.10266 [hep-th]}}].

\bibitem{Hellerman:2017veg}
S.~Hellerman, S.~Maeda and M.~Watanabe, {{Operator Dimensions from Moduli}},
  \href{http://dx.doi.org/10.1007/JHEP10(2017)089}{JHEP {\bf 10}, 089, 2017},
  [\href{http://arxiv.org/abs/arXiv:1706.05743}{{arXiv:1706.05743 [hep-th]}}].

\bibitem{dragnev2002discrete}
P.~D. Dragnev, D.~A. Legg and D.~W. Townsend, {Discrete logarithmic energy on
  the sphere}, {Pacific journal of mathematics {\bf 207}, 345--358, 2002}.

\bibitem{Bergersen_1994}
B.~Bergersen, D.~Boal and P.~Palffy-Muhoray 1994.

\bibitem{Badel:2019oxl}
G.~Badel, G.~Cuomo, A.~Monin and R.~Rattazzi, {{The Epsilon Expansion Meets
  Semiclassics}}, \href{http://dx.doi.org/10.1007/JHEP11(2019)110}{JHEP {\bf
  11}, 110, 2019},
  [\href{http://arxiv.org/abs/arXiv:1909.01269}{{arXiv:1909.01269 [hep-th]}}].

\bibitem{Hasebe:2010vp}
K.~Hasebe, {{Hopf Maps, Lowest Landau Level, and Fuzzy Spheres}},
  \href{http://dx.doi.org/10.3842/SIGMA.2010.071}{SIGMA {\bf 6}, 071, 2010},
  [\href{http://arxiv.org/abs/arXiv:1009.1192}{{arXiv:1009.1192 [hep-th]}}].

\bibitem{Lieb1973}
E.~H. Lieb, {The classical limit of quantum spin systems},
  \href{http://dx.doi.org/10.1007/BF01646493}{Communications in Mathematical
  Physics {\bf 31}, 327--340, 1973}.

\bibitem{Costa:2011mg}
M.~S. Costa, J.~Penedones, D.~Poland and S.~Rychkov, {{Spinning Conformal
  Correlators}}, \href{http://dx.doi.org/10.1007/JHEP11(2011)071}{JHEP {\bf
  11}, 071, 2011},
  [\href{http://arxiv.org/abs/arXiv:1107.3554}{{arXiv:1107.3554 [hep-th]}}].

\bibitem{Saff1997DistributingMP}
E.~Saff and A.~B.~J. Kuijlaars, {Distributing many points on a sphere}, {The
  Mathematical Intelligencer {\bf 19}, 5--11, 1997}.

\bibitem{tkachenko1966vortex}
V.~Tkachenko, {On vortex lattices}, {Sov. Phys. JETP {\bf 22}, 1282--1286,
  1966}.

\bibitem{PhysRevLett.43.214}
E.~J. Yarmchuk, M.~J.~V. Gordon and R.~E. Packard, {Observation of stationary
  vortex arrays in rotating superfluid helium},
  \href{http://dx.doi.org/10.1103/PhysRevLett.43.214}{Phys. Rev. Lett. {\bf
  43}, 214--217, 1979}.

\bibitem{Moroz:2018noc}
S.~Moroz, C.~Hoyos, C.~Benzoni and D.~T. Son, {{Effective field theory of a
  vortex lattice in a bosonic superfluid}},
  \href{http://dx.doi.org/10.21468/SciPostPhys.5.4.039}{SciPost Phys. {\bf 5},
  039, 2018}, [\href{http://arxiv.org/abs/arXiv:1803.10934}{{arXiv:1803.10934
  [cond-mat.quant-gas]}}].

\bibitem{Esposito:2017qpj}
A.~Esposito, S.~Garcia-Saenz, A.~Nicolis and R.~Penco, {{Conformal solids and
  holography}}, \href{http://dx.doi.org/10.1007/JHEP12(2017)113}{JHEP {\bf 12},
  113, 2017}, [\href{http://arxiv.org/abs/arXiv:1708.09391}{{arXiv:1708.09391
  [hep-th]}}].

\bibitem{tkachenko}
V.~Tkachenko, {Elasticity of vortex lattices}, {Soviet J. Exp. Theor. Phys {\bf
  29}, 945, 1969}.

\bibitem{PhysRev.123.1242}
W.~Kohn, {Cyclotron resonance and de haas-van alphen oscillations of an
  interacting electron gas},
  \href{http://dx.doi.org/10.1103/PhysRev.123.1242}{Phys. Rev. {\bf 123},
  1242--1244, 1961}.

\bibitem{delaFuente:2018qwv}
A.~De~La~Fuente, {{The large charge expansion at large $N$}},
  \href{http://dx.doi.org/10.1007/JHEP08(2018)041}{JHEP {\bf 08}, 041, 2018},
  [\href{http://arxiv.org/abs/arXiv:1805.00501}{{arXiv:1805.00501 [hep-th]}}].

\bibitem{Alvarez-Gaume:2019biu}
L.~Alvarez-Gaume, D.~Orlando and S.~Reffert, {{Large charge at large N}},
  2019, [\href{http://arxiv.org/abs/arXiv:1909.02571}{{arXiv:1909.02571
  [hep-th]}}].

\bibitem{PhysRevB.48.13749}
W.~Chen, M.~P.~A. Fisher and Y.-S. Wu, {Mott transition in an anyon gas},
  \href{http://dx.doi.org/10.1103/PhysRevB.48.13749}{Phys. Rev. B {\bf 48},
  13749--13761, 1993}.

\bibitem{Seiberg:2016gmd}
N.~Seiberg, T.~Senthil, C.~Wang and E.~Witten, {{A Duality Web in 2+1
  Dimensions and Condensed Matter Physics}},
  \href{http://dx.doi.org/10.1016/j.aop.2016.08.007}{Annals Phys. {\bf 374},
  395--433, 2016},
  [\href{http://arxiv.org/abs/arXiv:1606.01989}{{arXiv:1606.01989 [hep-th]}}].

\bibitem{Karch:2016sxi}
A.~Karch and D.~Tong, {{Particle-Vortex Duality from 3d Bosonization}},
  \href{http://dx.doi.org/10.1103/PhysRevX.6.031043}{Phys. Rev. X {\bf 6},
  031043, 2016},
  [\href{http://arxiv.org/abs/arXiv:1606.01893}{{arXiv:1606.01893 [hep-th]}}].

\bibitem{PhysRevLett.65.2070}
P.~B. Wiegmann, {Parity violation and superconductivity in two-dimensional
  correlated electronic systems},
  \href{http://dx.doi.org/10.1103/PhysRevLett.65.2070}{Phys. Rev. Lett. {\bf
  65}, 2070--2073, 1990}.

\bibitem{PhysRevB.95.085151}
D.~X. Nguyen and A.~Gromov, {{Exact Electromagnetic Response of Landau Level
  Electrons}}, \href{http://dx.doi.org/10.1103/PhysRevB.95.085151}{Phys. Rev. B
  {\bf 95}, 085151, 2017},
  [\href{http://arxiv.org/abs/arXiv:1610.03516}{{arXiv:1610.03516
  [cond-mat.str-el]}}].

\bibitem{Du:2020gqf}
Y.-H. Du, U.~Mehta and D.~T. Son, {{Rotons in Anyon Superfluids}},  2020,
  [\href{http://arxiv.org/abs/arXiv:2012.07991}{{arXiv:2012.07991
  [cond-mat.mes-hall]}}].

\bibitem{Grassi:2019txd}
A.~Grassi, Z.~Komargodski and L.~Tizzano, {{Extremal Correlators and Random
  Matrix Theory}},  2019,
  [\href{http://arxiv.org/abs/arXiv:1908.10306}{{arXiv:1908.10306 [hep-th]}}].

\bibitem{Watanabe:2019adh}
M.~Watanabe, {{Chern-Simons-Matter Theories at Large Global Charge}},  2019,
  [\href{http://arxiv.org/abs/arXiv:1904.09815}{{arXiv:1904.09815 [hep-th]}}].

\bibitem{Chester:2017vdh}
S.~M. Chester, L.~V. Iliesiu, M.~Mezei and S.~S. Pufu, {{Monopole Operators in
  $U(1)$ Chern-Simons-Matter Theories}},
  \href{http://dx.doi.org/10.1007/JHEP05(2018)157}{JHEP {\bf 05}, 157, 2018},
  [\href{http://arxiv.org/abs/arXiv:1710.00654}{{arXiv:1710.00654 [hep-th]}}].

\bibitem{Radicevic:2015yla}
D.~Radi\v{c}evi\'c, {{Disorder Operators in Chern-Simons-Fermion Theories}},
  \href{http://dx.doi.org/10.1007/JHEP03(2016)131}{JHEP {\bf 03}, 131, 2016},
  [\href{http://arxiv.org/abs/arXiv:1511.01902}{{arXiv:1511.01902 [hep-th]}}].

\bibitem{Geracie:2015drf}
M.~Geracie, M.~Goykhman and D.~T. Son, {{Dense Chern-Simons Matter with
  Fermions at Large N}}, \href{http://dx.doi.org/10.1007/JHEP04(2016)103}{JHEP
  {\bf 04}, 103, 2016},
  [\href{http://arxiv.org/abs/arXiv:1511.04772}{{arXiv:1511.04772 [hep-th]}}].

\bibitem{Minwalla:2020ysu}
S.~Minwalla, A.~Mishra and N.~Prabhakar, {{Fermi seas from Bose condensates in
  Chern-Simons matter theories and a bosonic exclusion principle}},  2020,
  [\href{http://arxiv.org/abs/arXiv:2008.00024}{{arXiv:2008.00024 [hep-th]}}].

\bibitem{Berkooz:2006wc}
M.~Berkooz, D.~Reichmann and J.~Simon, {{A Fermi Surface Model for Large
  Supersymmetric AdS(5) Black Holes}},
  \href{http://dx.doi.org/10.1088/1126-6708/2007/01/048}{JHEP {\bf 01}, 048,
  2007},
  [\href{http://arxiv.org/abs/arXiv:hep-th/0604023}{{arXiv:hep-th/0604023}}].

\bibitem{Berkooz:2008gc}
M.~Berkooz and D.~Reichmann, {{Weakly Renormalized Near 1/16 SUSY Fermi Liquid
  Operators in N=4 SYM}},
  \href{http://dx.doi.org/10.1088/1126-6708/2008/10/084}{JHEP {\bf 10}, 084,
  2008}, [\href{http://arxiv.org/abs/arXiv:0807.0559}{{arXiv:0807.0559
  [hep-th]}}].

\bibitem{Lee:2009epi}
S.-S. Lee, {{Low energy effective theory of Fermi surface coupled with U(1)
  gauge field in 2+1 dimensions}},
  \href{http://dx.doi.org/10.1103/PhysRevB.80.165102}{Phys. Rev. B {\bf 80},
  165102, 2009}, [\href{http://arxiv.org/abs/arXiv:0905.4532}{{arXiv:0905.4532
  [cond-mat.str-el]}}].

\bibitem{Giombi:2011kc}
S.~Giombi, S.~Minwalla, S.~Prakash, S.~P. Trivedi, S.~R. Wadia and X.~Yin,
  {{Chern-Simons Theory with Vector Fermion Matter}},
  \href{http://dx.doi.org/10.1140/epjc/s10052-012-2112-0}{Eur. Phys. J. C {\bf
  72}, 2112, 2012},
  [\href{http://arxiv.org/abs/arXiv:1110.4386}{{arXiv:1110.4386 [hep-th]}}].

\bibitem{Damia:2019bdx}
J.~Aguilera~Damia, S.~Kachru, S.~Raghu and G.~Torroba, {{Two dimensional
  non-Fermi liquid metals: a solvable large N limit}},
  \href{http://dx.doi.org/10.1103/PhysRevLett.123.096402}{Phys. Rev. Lett. {\bf
  123}, 096402, 2019},
  [\href{http://arxiv.org/abs/arXiv:1905.08256}{{arXiv:1905.08256
  [cond-mat.str-el]}}].

\bibitem{Kravec:2018qnu}
S.~Kravec and S.~Pal, {{Nonrelativistic Conformal Field Theories in the Large
  Charge Sector}}, \href{http://dx.doi.org/10.1007/JHEP02(2019)008}{JHEP {\bf
  02}, 008, 2019},
  [\href{http://arxiv.org/abs/arXiv:1809.08188}{{arXiv:1809.08188 [hep-th]}}].

\bibitem{Hoyos:2013eha}
C.~Hoyos, S.~Moroz and D.~T. Son, {{Effective theory of chiral two-dimensional
  superfluids}}, \href{http://dx.doi.org/10.1103/PhysRevB.89.174507}{Phys. Rev.
  B {\bf 89}, 174507, 2014},
  [\href{http://arxiv.org/abs/arXiv:1305.3925}{{arXiv:1305.3925
  [cond-mat.quant-gas]}}].

\end{thebibliography}\endgroup

\end{document}